\documentclass[10pt,a4paper]{article}

\pdfoutput=1        
\usepackage[utf8]{inputenc}
\usepackage{amsmath}
\usepackage{amsfonts}
\usepackage{amssymb}
\usepackage{dsfont}
\usepackage{upgreek}
\usepackage{physics}
\usepackage{graphicx}

\usepackage{fancyhdr}   

\usepackage{cite}

\fancyheadoffset{0pt}

\numberwithin{equation}{section}
\numberwithin{figure}{section}
\numberwithin{table}{section}

\author{
  Riccardo Finotello\thanks{E-mail: riccardo.finotello@to.infn.it} \and
  Igor Pesando\thanks{E-mail: ipesando@to.infn.it} \\[0.75cm]
  Dipartimento di Fisica, Universit\`{a} di Torino \\
  and I.N.F.N. - sezione di Torino \\[0.3cm]
  Via P. Giuria 1, I-10125 Torino, Italy \\
}

\title{2D Fermion on the Strip with Boundary Defects as a CFT with
  Excited Spin Fields}
\date{\today}

\newcommand{\myoverline}[1]{#1}                    

\newcommand{\Nf}{\mathrm{N}_f}                               
\newcommand{\SM}{S_{\mathcal{M}}}                            
\newcommand{\SE}{S_E}                                        
\newcommand{\uppsibar}{\overline{\uppsi}}                    
\newcommand{\huppsi}{{\widehat{\uppsi}}}                     
\newcommand{\hUppsi}{{\widehat{\Uppsi}}}                     
\newcommand{\lrpartial}[1]{\overset{
                             \leftrightarrow
                           }{
                             \partial_{ #1 }
                           }
                          }                                  
\newcommand{\R}[1]{\mathrm{R}_{\qty( #1 )}}                  
\newcommand{\uptauhat}{\hat{\uptau}}                         
\newcommand{\UN}{\mathrm{U}( \mathrm{N}_f )}                 
\newcommand{\consprod}[2]
  {\left\langle #1, #2 \right\rangle}                        
\newcommand{\lconsprod}[2]
  {\left\langle\hspace{-0.25em}\left\langle #1\right.,
  #2 \right\rangle}                                          
\newcommand{\dual}[1]{{}^*#1}                                
\newcommand{\En}{\mathrm{H}}                                 
\newcommand{\bupxi}{\overline{\upxi}}                        
\newcommand{\bupzeta}{\overline{\upzeta}}                    
\newcommand{\Lvir}[1]{\mathrm{L}_{#1}}                       
\newcommand{\bu}{\overline{u}}                               
\newcommand{\bz}{\overline{z}}                               
\newcommand{\no}[1]{\;: #1 :}                                
\newcommand{\noE}[1]{\;N_{E,\bar E} \left[ #1 \right]}       
\newcommand{\sumt}{\sum\limits_{t=1}^N}                      
\newcommand{\prodt}{\prod\limits_{t=1}^N}                    
\newcommand{\eps}[1]{\upepsilon_{\qty( #1 )}}                
\newcommand{\epss}{\upepsilon}                               
\newcommand{\beps}[1]{\overline{\upepsilon}_{\qty( #1 )}}    
\newcommand{\bepss}{\overline{\upepsilon}}                   
\newcommand{\normfac}{\mathcal{N}_{\Uppsi}}                  
\newcommand{\E}[1]{\mathrm{E}_{\qty( #1 )}}                  
\newcommand{\tE}[1]{\widetilde{\mathrm{E}}_{\qty( #1 )}}     
\newcommand{\Es}{\mathrm{E}}                                 
\newcommand{\tEs}{\widetilde{\mathrm{E}}}
\newcommand{\bE}[1]{\overline{\mathrm{E}}_{\qty( #1 )}}      
\newcommand{\btE}[1]{                        \overline{
                     \widetilde{
    \mathrm{E}
                        }
                      }_{\qty( #1 )}
                    }                                        
\newcommand{\bEs}{\overline{\mathrm{E}}}                     
\newcommand{\btEs}{\overline{\widetilde{\mathrm{E}}}}
\newcommand{\obE}{ b^{( \Es )} }
\newcommand{\obbE}{ b^{*\, ( \bEs )} }
\newcommand{\LL}[1]{\mathrm{L}_{\qty( #1 )}}                 
\newcommand{\LLs}{\mathrm{L}}                                
\newcommand{\M}{\mathrm{M}}                                  
\newcommand{\bM}{\overline{\mathrm{M}}}                      

\newcommand{\Gexcvac}{ { \Omega_{\{x_t, \E t,\bE t\}} }}
\newcommand{\Gexcvacket}{{ \ket{\Gexcvac} }}

\newcommand{\GGexcvac}{ { \Omega_{\{x_t, \E t\}} }}
\newcommand{\GGexcvacket}{{ \ket{\GGexcvac} }}
\newcommand{\GGexcvacbra}{{\bra{\GGexcvac} }}

\newcommand{\excvacket}{\ket{T_{\Es,\bEs}}}                  

\newcommand{\eexcvacket}{\ket{T_{\Es}}}                  
\newcommand{\eexcvacbra}{\bra{T_{\Es}}}                  

\newcommand{\twsvacket}{\ket{\mathrm{T}}}                    
\newcommand{\SL}{ { \mathrm{SL}_2\qty( \mathds{R} ) } }            
\newcommand{\cmode}[4]{\mathfrak{C}_{ #1 }
                       \qty(
                          #2,
                          \qty{ #4 , #3 }
                       )
                      }                                      
\newcommand{\spin}[2]{\mathrm{S}_{ #1 }\qty( #2 )}           

\begin{document}

  \maketitle

  \begin{abstract}
    We consider a two-dimensional fermion on the strip in the presence
of an arbitrary number of zero-dimensional boundary changing defects.
We show that the theory is still conformal with time dependent stress-energy
tensor and that the allowed defects can be
understood as excited spin fields.  Finally we compute  correlation
functions involving these excited spin fields without using bosonization.
  \end{abstract}

  \clearpage

  \tableofcontents

  \clearpage

  \section{Introduction and Conclusion}

  The study of viable phenomenological models in the framework of String
  Theory often involves the analysis of the properties of systems of
  D-branes. Clearly the inclusion of the physical requirements needed for a
  consistent theory deeply constrains the possible scenarios. In particular the
  chiral spectrum of the Standard Model acts as a strong restriction on the
  possible D-brane setup. Intersecting and magnetized branes
  represent  relevant classes of such models with interacting chiral
  matter. 
  In particular most models involve a compactification with
  factorized two-tori and magnetic backgrounds
  (see \cite{Blumenhagen:2005mu,Ibanez:2012zz} for review)
  and only few attempts
  have been done to study more general cases. 

  The computation of interesting quantities such as Yukawa couplings
  involves quite often
  correlators of excited spin and twist fields.
  The correlators of (excited) spin  fields has been a research subject
  for many years until the formulation found in the seminal paper by
  Friedan, Martinec and Shenker \cite{Friedan:1985ge}
  based on bosonization.  
  On the other side the correlators of excited twist fields has not been
  algorithmic until recent \cite{Pesando:2014owa,Pesando:2011ce}
  (see for example
  \cite{Erler:1992gt,Anastasopoulos:2011gn,Anastasopoulos:2011hj,Anastasopoulos:2013sta}
  for earlier work on excited twist fields
  and
  \cite{Hamidi:1986vh,Harvey:1986bf,Atick:1987kd,Goodman:1988av,Abel:2003vv,Cvetic:2003ch,Abel:2003yx,Pesando:2009tt,Pesando:2012cx}
  for the basic correlators).
 Even for the spin field case the available techniques allow to compute
 only correlators involving ``Abelian'' configurations, i.e. configurations
 which can be factorized in sub-configurations having $\mathrm{U}(1)$ symmetry
 and
 only few papers have considered the non Abelian case
 \cite{Inoue:1987ak,Inoue:1990ci,Gato:1990mx,Frampton:2000mq,Pesando:2015fpj,Finotello:2018bhn
   } which is mathematically by far more complicated and related to
   the unresolved connection problem of Fuchsian equations.
  
  Despite the existence of an efficient method based on bosonization
  \cite{Friedan:1985ge} for computing correlators involving excited
  spin fields we re-examine the problem and give 
  a new method to compute such correlators
  which adds on the present one and
  the very old one based on Reggeon vertex
  \cite{Sciuto:1969vz,DellaSelva:1970bj,Schwarz:1973jf,DiVecchia:1989hf,Nilsson:1989fz,DiBartolomeo:1990fw,Engberg:1992sk,Petersen:1988gj}.

  One reason for such a research
  is that we hope to be able to extend this approach to
  correlators involving twist fields
  and non Abelian spin and twist fields.
  In particular we would also like to clarify the reason of
  the non existence of an approach equivalent to bosonization for
  twist fields.

  Another reason is that we are interested to explore what happens to
  a CFT in presence of defects.
  It turns out that despite the defects it is still possible to define
  a radial time dependent stress-energy tensor which satisfies the
  canonical OPE with the right central charge.
  Moreover the  boundary changing defects in the
  construction  can be
  associated with excited spin fields
  and this allows to compute correlators involving excited spin fields
  without resorting to bosonization.

The paper is organized as follows.
  In Section~\ref{sec:Mink_theory} we define the Minkowskian
  formulation of the theory we are interested in and we introduce the notation.
  Then in Section~\ref{sec:product} we discuss the conserved quantities.
  In particular we introduce a conserved product
  used to extract the coefficients of the expansion of the
  fields in modes.
  In order to obtain these coefficients, which we want to interpret as creation and
  annihilation operators, we are led to the introduction of the space of
  dual modes.
  
  In Section~\ref{sec:eucl_formulation} we discuss the Euclidean
  formulation on the strip and the upper plane not relying on the CFT
  properties since we have not yet shown that the theory is a CFT.

  In Section~\ref{sec:modes_and_algebra}
  we find the explicit expression of the
  modes which satisfy the equations of motion and the boundary conditions.
  Then we compute the dual modes and finally the algebra of the
  creators and annihilators.
  This step is conceptually separated from the definition of the in-vacuum and
  the Fock space where this algebra is represented. We take care of this in
  Section~\ref{sec:invacuum}.

  In Section~\ref{sect:asymp_fields} we relate the fermionic field
  with its asymptotic in- and out- counterparts.
  This is useful in the last section in order to justify the new way
  of computing the excited spin fields correlators.
  
  With the definition of the vacuum we have an  associated normal ordering.
  In Section~\ref{sec:contraction_and_T} we compute
  the contractions and OPEs of the operators and define the stress-energy
  tensor which satisfies the canonical CFT algebra, thus showing that
  the theory is a CFT.
  Then we argue that the defects are excited spin fields.
 
  In Section~\ref{sec:hermitian_and_outvacuum} we take care of the
  definition of the operation which we want to interpret as Euclidean
  Hermitian conjugation when we define the bra of the vacuum.
  This operation is almost the same as the $\star$ operator defined in
  the algebraic approach to QFT.
  Using this definition, which we want to be the Hermitian conjugation,
  in Section~\ref{sec:out-vacuum} we define the bra-vacuum as it is
  necessary to compute the correlators.

  Finally in Section~\ref{sec:spin_correlators}
  we compute correlators of excited spin fields
  using the in- and out-vacua in presence of defects.

  \section{Point-like Defect CFT: the Minkowskian Formulation}
  \label{sec:Mink_theory}
  
  In this section we introduce the theory we study by presenting its
  worldsheet action and boundary conditions in the presence of $N$
  zero-dimensional defects. This theory is later shown to be a CFT
  despite the existence of defects.

    \subsection{Action Principle and Boundary Conditions}

    Let $\qty( \uptau, \upsigma ) \in \Upsigma = \qty( -\infty, +\infty) \times
    \qty[ 0, \uppi ]$ define a strip with Lorentzian metric\footnote{
      We consider the metric 
      \begin{equation*}
        \dd{s}^2 = -\dd{\uptau}^2 + \dd{\upsigma}^2,
      \end{equation*}
      and the lightcone coordinates
      $\upxi_{\pm} = \uptau \pm \upsigma$ which allow to define
      \begin{equation*}
      \partial_{\pm} = \frac{1}{2} \qty( \partial_{\uptau} \pm
      \partial_{\upsigma} )
      \end{equation*}
      and
      \begin{equation*}
        \dd^2{\upxi}=\frac{1}{2}\dd{\upxi_+} \dd{\upxi_-} =\dd{\uptau}
        \dd{\upsigma}.
      \end{equation*}
      The anti-symmetric tensor is $\upepsilon_{\uptau \upsigma} =
      -\upepsilon^{\uptau \upsigma} = 1$ and the gamma matrices are
      \begin{equation*}
        \upgamma^{\uptau} = \mqty( \admat{1,-1} ) = -\upgamma_{\uptau},
        \qquad
        \upgamma^{\upsigma} = \mqty( \admat{1,1} ) = \upgamma_{\upsigma}.
      \end{equation*}
      
      We also consider the two-dimensional spinor
      \begin{equation*}
        \uppsi = \mqty( \uppsi_+ \\ \uppsi_- ),
      \end{equation*}
      whose conjugate is $\uppsibar = \uppsi^{\dagger} \upgamma^{\uptau} =
      \mqty( -\uppsi_-^* & \uppsi_+^* )$.
    }
    and consider $\Nf$ massless complex fermions $\uppsi^i$ such that $i = 1, 2,
    \dots, \Nf$. Their two-dimensional Minkowski action defined on the strip
    $\Upsigma$ is:
    \begin{equation*}
      \SM = \frac{T}{2} \int\limits_{-\infty}^{+\infty} \dd{\uptau}
      \int\limits_0^{+\uppi} \dd{\upsigma}
      \qty( \frac{1}{2} \uppsibar_i( \uptau, \upsigma )
      \qty( -i \upgamma^{\upalpha} \lrpartial{\upalpha} )
      \uppsi^i( \uptau, \upsigma ) ).
      \label{eq:cft-action_full}
    \end{equation*}
    In components the action reads:
    \begin{equation}
      \SM = i \frac{T}{2} \iint \dd^2{\upxi} \qty( \uppsi^*_{-,\, i} \lrpartial{+}
      \uppsi^i_- + \uppsi^*_{+,\, i} \lrpartial{-} \uppsi^i_+ ) ,
      \label{eq:cft-action}
    \end{equation}
    so the equations of motion are:
    \begin{equation}
      \begin{split}
        \partial_- \uppsi_{+}^i( \upxi_+, \upxi_- ) & = \partial_+
        \uppsi_{-}^i( \upxi_+, \upxi_- ) = 0,
        \\
        \partial_- \uppsi^*_{+,\, i}( \upxi_+, \upxi_- ) & = \partial_+
        \uppsi^*_{-,\, i}( \upxi_+, \upxi_- ) = 0.
      \end{split}
      \label{eq:eom}
    \end{equation}
    Their solutions are the usual ``holomorphic'' functions $\uppsi_{+}^i(
    \upxi_+)$ and $\uppsi_{-}^i( \upxi_- )$, together with their complex
    conjugates\footnote{
      Notice that $\uppsi^*$ is indeed the complex conjugate of the field
      $\uppsi$, while it will no longer be the case in the Euclidean formalism.
    }.

    \begin{figure}[t]
      \centering
      \includegraphics{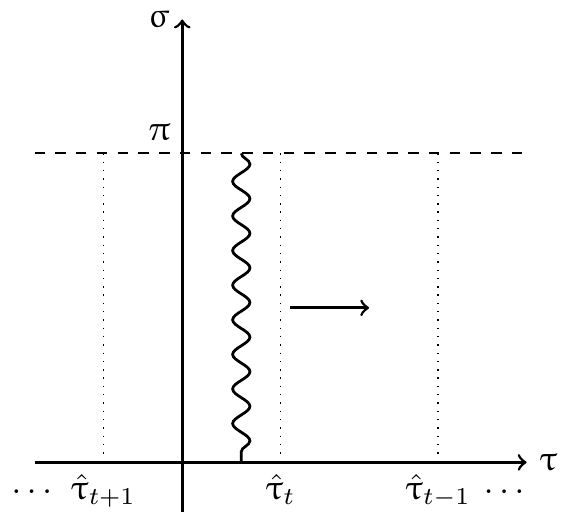}
      \caption{We describe the propagation of a string in the presence of
      point-like defects in the time direction. Each point $\uptauhat_t$ on the
      boundary of the strip is in correspondence to a non trivial change in the
      boundary conditions.}
      \label{fig:point-like-defects}
    \end{figure}

    The boundary conditions are instead:
    \begin{equation}
      \eval{
        \qty(
          \var{\uppsi}_{+,\, i}^* \uppsi_{+}^{ i} +
          \var{\uppsi}_{-,\, i}^* \uppsi_{-}^{ i} -
          \uppsi_{+,\, i}^* \var{\uppsi}_{+}^{ i} -
          \uppsi_{-,\, i}^* \var{\uppsi}_{-}^{ i}
        )
      }_{\upsigma = 0}^{\upsigma = \uppi} = 0.
      \label{eq:boundary-conditions}
    \end{equation}
    We solve the constraint imposing the non trivial relations:
    \begin{equation}
      \begin{cases}
        \uppsi_-^i( \uptau, 0 ) = \qty( \R{t} )^i_j \uppsi^j_+( \uptau, 0
        ) & \qfor \uptau \in \qty( \uptauhat_t, \uptauhat_{t-1} ),
        \\
        \uppsi_-^i( \uptau, \uppi ) = - \uppsi_+^i( \uptau, \uppi ) & \qfor
        \uptau \in \mathds{R},
      \end{cases}
      \label{eq:boundary-conditions-solutions}
    \end{equation}
    where $t = 1, 2, \dots, N$.
    This way we introduced $N$ zero-dimensional defects on the boundary
    as in Figure~\ref{fig:point-like-defects}.  They are located on the strip at
    $( \uptauhat_t, 0 )$ such that $\uptauhat_t < \uptauhat_{t-1}$ with
    $\uptauhat_{N+1} = -\infty$ and $\uptauhat_0 = +\infty$. We have also
    introduced $N$ matrices $\R{t} \in \UN$ which characterize the defects.
    
    Since in most of this paper we want the in- and out-vacua to be the usual NS
    vacuum, we have chosen the boundary condition at $\upsigma = \uppi$ so that
    when there are no defects the system describes NS fermions. We require also
    the cancellation of the action of the defects at $\uptauhat = \pm\infty$,
    i.e.:
    \begin{equation*}
      \R{N} \R{N-1} \dots \R{1} = \mathds{1}.
    \end{equation*}
    More general cases where in- and/or out-vacua are twisted can be worked out
    similarly to what we do.
   
    In order to connect to the Euclidean formulation we introduce $\Nf$
    ``double fields''\footnote{In this case they correspond to the fields
    $\uppsi^i_+$.} $\Uppsi^i$ obtained by gluing $\uppsi^i_+$ and $\uppsi^i_-$
    along the $\upsigma=\uppi$ boundary and labeled by an index $i = 1, 2,
    \dots, \Nf$:
    \begin{equation}
      \Uppsi^i(\uptau, \upphi)=
      \begin{cases}
        \uppsi^i_+(\uptau, \upphi) &\qfor  0\le\upphi\le \uppi \\
        -\uppsi^i_-(\uptau, 2\uppi-\upphi) &\qfor \uppi\le\upphi\le 2\uppi        
      \end{cases}
%
      \label{eq:double-field-Lorentzian}
    \end{equation}
    where $0\le\upphi\le 2\uppi        $.
    The boundary conditions become:
    \begin{equation*}
      \Uppsi^i(\uptau, 2 \uppi ) =
      - \qty( \R{t} )^i_j \Uppsi^j(\uptau, 0 ),\quad
      \uptau
      \in \qty( \uptauhat_t, \uptauhat_{t-1} ).
    \end{equation*}
    Using the equation of motion we get
    $\Uppsi^i(\uptau, \upphi)=\Uppsi^i(\uptau+ \upphi)$
    and the boundary conditions become the (pseudo)periodicity
    conditions
    \begin{equation*}
      \Uppsi^i(\uptau + 2 \uppi ) =
      - \qty( \R{t} )^i_j \Uppsi^j(\uptau ),\quad
      \uptau
      \in \qty( \uptauhat_t, \uptauhat_{t-1} ).
    \end{equation*}
    We will use them to write some expressions similar to the
    Euclidean ones.

    The main issue is now to expand $\Uppsi$ in a basis of modes and proceed to
    its quantization. Even in the simplest case $\Nf = 1$ the task of finding
    the Minkowskian modes turns out to be fairly complicated. It is however
    possible to overcome the issue in the Euclidean formalism.

  \section{Conserved Product and Charges}
\label{sec:product}
  In order to have a good quantum formulation, we define a procedure to
  build a Fock space of states in the Heisenberg formalism thus equal time
  anti-commutation relations must be invariant in time. We therefore need
  a time independent internal product to extract the creation and annihilation
  operators and expand the fields on the basis of modes.

    \subsection{Definition of the Conserved Product}

    Start from a generic conserved current
    \begin{equation*}
      j( \uptau, \upsigma ) = j_{\uptau}( \uptau, \upsigma ) \dd{\uptau} +
      j_{\upsigma}( \uptau, \upsigma ) \dd{\upsigma},
    \end{equation*}
    and consider
    \begin{equation*}
      \star j = j_{\upsigma} \dd{\uptau} + j_{\uptau} \dd{\upsigma} \quad
      \Rightarrow \quad \dd{(\star j)} = \qty( \partial_{\uptau} j_{\uptau} -
      \partial_{\upsigma} j_{\upsigma} ) \dd{\uptau} \dd{\upsigma},
    \end{equation*}
    where $\star$ is the Hodge dual operator. Integrating the 2-form over a
    surface $\Upsigma' = \qty[ \uptau_i, \uptau_f ] \times \qty[ 0, \uppi ]$
    yields:
    \begin{equation*}
      \begin{split}
        \int\limits_{\Upsigma'} \dd{(\star j)} = \int\limits_{\partial\Upsigma'}
        \star j & = 0 \Leftrightarrow
        \\
        \Leftrightarrow \int\limits_0^{\uppi} \dd{\upsigma} \qty(
        \eval{j_{\uptau}}_{\uptau = \uptau_f} - \eval{j_{\uptau}}_{\uptau =
        \uptau_i} ) & = \int\limits_{\uptau_i}^{\uptau_f} \dd{\uptau} \qty(
        \eval{j_{\upsigma}}_{\upsigma = \uppi} - \eval{j_{\upsigma}}_{\upsigma =
        0} ).
      \end{split}
    \end{equation*}
    The current $j_{\uptau}( \uptau, \upsigma )$ is thus conserved in time if
    \begin{equation}
      \int\limits_{\uptau_i}^{\uptau_f} \dd{\uptau} \qty(
      \eval{j_{\upsigma}}_{\upsigma = \uppi} - \eval{j_{\upsigma}}_{\upsigma =
      0} ) = 0.
      \label{eq:time-conservation}
    \end{equation}
    If this is the case, then we can say that
    \begin{equation*}
      Q = \int\limits_0^{\uppi} \dd{\upsigma} j_{\uptau}( \uptau,
      \upsigma ) 
    \end{equation*}
    is conserved (that is $\partial_{\uptau} Q = 0$).

    We now consider explicitly the symmetries of the action
    \eqref{eq:cft-action}. In particular we focus on the diffeomorphism
    invariance and $\UN$ flavour symmetries of the bulk theory leading to the
    stress-energy tensor and a vector current. We apply the aforementioned
    procedure to these objects to study their properties and (non-)conservation.

      \subsubsection{Flavour Vector Current}

      Consider first the $\UN$ vector current generated by the flavour symmetry
      of the action \eqref{eq:cft-action_full}. In general we can write it as
      \begin{equation*}
        j_{\upalpha}^a ( \uptau, \upsigma ) = \qty( \mathrm{T}^a )^i_j
        \uppsibar_i( \uptau, \upsigma ) \upgamma_{\upalpha} \uppsi^j( \uptau,
        \upsigma ),
      \end{equation*}
      where $\mathrm{T}^a$ is in principle a generator of $\UN$ ($a = 1, 2,
      \dots, \Nf^2$), but the result holds for a generic matrix. The spinors
      $\uppsi$ and $\uppsibar$ can also be generalized to two different and
      arbitrary solutions to the equations of motion \eqref{eq:eom}
      while keeping the current conserved.
      In components we have:
      \begin{eqnarray*}
        j^a_{\uptau}( \uptau, \upsigma ) & = & \qty( \mathrm{T}^a )^i_j \qty(
        \uppsi^*_{+,\, i} \uppsi^j_+ + \uppsi^*_{-,\, i} \uppsi^j_- )
        \\
        j^a_{\upsigma}( \uptau, \upsigma ) & = & \qty( \mathrm{T}^a )^i_j
        \qty( \uppsi^*_{+,\, i} \uppsi^j_+ - \uppsi^*_{-,\, i} \uppsi^j_- ).
      \end{eqnarray*}

      In order to define a conserved charge, we require:
      \begin{equation*}
        \int\limits_{\uptau_i}^{\uptau_f} \dd{\uptau} \qty(
        \eval{j_{\upsigma}^a}_{\upsigma = \uppi} - \eval{
          j_{\upsigma}^a}_{\upsigma = 0} ) = 0,
      \end{equation*}
      where
      \begin{equation*}
        \eval{j_{\upsigma}^a( \uptau, \upsigma )}_{\upsigma = \uppi} \equiv 0
      \end{equation*}
      using the boundary conditions \eqref{eq:boundary-conditions}, and
      \begin{equation*}
        \eval{j_{\upsigma}^a( \uptau, \upsigma )}_{\upsigma = 0} = \qty[
          \uppsi^*_+ \qty( \mathrm{T}^a - \R{t}^{\dagger} \mathrm{T}^a \R{t} )
          \uppsi_+ ]_{\upsigma = 0},
        \quad
        \uptau \in \qty( \uptauhat_t, \uptauhat_{t-1} ).
      \end{equation*}
      In general
      \begin{equation*}
        \eval{j_{\upsigma}^a( \uptau, \upsigma )}_{\upsigma = 0} = 0
        \quad \Leftrightarrow \quad
        \mathrm{T}^a \propto \mathds{1}
      \end{equation*}
      so that $\R{t}^{\dagger} \mathrm{T}^a = \mathrm{T}^a \R{t}^{\dagger}$.
      This shows that the presence of the point-like defects on the worldsheet
      generally breaks the $\UN$ symmetry ($\mathrm{SO}( \Nf ) \times
      \mathrm{SO}( \Nf )$ if we consider Majorana-Weyl fermions) down to a
      $\mathrm{U}( 1 )$ phase because of the boundary conditions
      \eqref{eq:boundary-conditions}. The $\mathrm{U}( 1 )$ vector current then
      defines a conserved charge for a restricted class of functions.

      Let $\upalpha$ and $\upbeta$ be two arbitrary (bosonic)
      solutions to the equations
      of motion \eqref{eq:eom}, we can in fact define a product
      \begin{equation}
        \consprod{\upalpha}{\upbeta} = \mathcal{N} \int\limits_0^{\uppi}
        \dd{\upsigma} \qty( \upalpha_{+,\, i}^* \upbeta_+^i + \upalpha_{-,\, i}^*
        \upbeta_-^i ),
        \label{eq:conserved-product}
      \end{equation}
      where $\mathcal{N} \in \mathds{R}$ is a normalization constant and the
      integrand must be free of non integrable singularities. The product is
      such that
      \begin{equation*}
        \consprod{\upalpha}{\upbeta} = \consprod{\upalpha}{\upbeta}^*.
      \end{equation*}

      We can also rewrite the result to the double fields defined in
      \eqref{eq:double-field-Lorentzian}. Let $A$ and $B$ be the ``double
      fields'' corresponding to $\upalpha$ and $\upbeta$ respectively, then we
      have:
      \begin{equation}
        \consprod{\upalpha}{\upbeta} = \mathcal{N} \int\limits_{0}^{2 \uppi}
        \dd{\upphi} A_i^*( \uptau + \upphi ) B^i( \uptau + \upphi ).
        \label{eq:conserved-product-double-field}
      \end{equation}

      \subsubsection{Stress-Energy Tensor}

      We now consider the stress-energy tensor of the bulk theory.
      The Noether procedure gives the off-shell tensor components
      \begin{equation*}
        \begin{split}
          \mathcal{T}_{\pm\pm} = &
          -i \frac{T}{4} \uppsi_{\pm,\, i}^*
          \lrpartial{\pm} \uppsi_+^i,
          \\
          \mathcal{T}_{\pm\mp} = &
          +i \frac{T}{4} \uppsi_{\mp,\, i}^*
          \lrpartial{\pm} \uppsi_\mp^i
          ,
        \end{split}
      \end{equation*}
      which become
      \begin{equation}
        \begin{split}
          \mathcal{T}_{++}( \upxi_+ ) = & -i \frac{T}{4} \uppsi_{+,\, i}^*(
          \upxi_+) \lrpartial{+} \uppsi_+^i( \upxi_+ ),
          \\
          \mathcal{T}_{--}( \upxi_- ) = & -i \frac{T}{4} \uppsi_{-,\, i}^*(
          \upxi_-)\lrpartial{-} \uppsi_-^i( \upxi_- )
        \end{split}
        \label{eq:stress-energy-tensor-lightcone}
      \end{equation}
      when on-shell. The boundary breaks the symmetry for translations in the
      $\upsigma$ direction, while the defects break the time translations: the
      Hamiltonian is therefore time-dependent but it is constant between two
      consecutive point-like defects.

      From the definition of the stress-energy tensor we can in principle
      build the hypothetical charges:
      \begin{eqnarray}
        \En( \uptau ) & = & \int\limits_0^{\uppi} \dd{\upsigma}
        \mathcal{T}_{\uptau\uptau}( \uptau, \upsigma ) = \int\limits_0^{\uppi}
        \dd{\upsigma} \qty( \mathcal{T}_{++}( \uptau + \upsigma ) +
        \mathcal{T}_{--}( \uptau - \upsigma ) ),
        \label{eq:hamiltonian}
        \\
        \mathrm{P}( \uptau ) & = & \int\limits_0^{\uppi} \dd{\upsigma}
        \mathcal{T}_{\uptau\upsigma}( \uptau, \upsigma ) = \int\limits_0^{\uppi}
        \dd{\upsigma} \qty( \mathcal{T}_{++}( \uptau + \upsigma ) -
        \mathcal{T}_{--}( \uptau - \upsigma ) ),
        \label{eq:momentum}
      \end{eqnarray}
      which are conserved if \eqref{eq:time-conservation} holds. Let the
      point-like defects be ordered as $\uptauhat_{t_0 - 1} < \uptau_i \le
      \uptauhat_{t_0} < \uptauhat_{t_N} \le \uptau_f < \uptauhat_{t_N+1}$, then
      for the linear momentum $\mathrm{P}$ the condition of conservation
      reads\footnote{
        Notice that in the second term of the second line, the differentiation
        with respect to $\uptau$ is acting only on $\R{t}$ and
        $\R{t}^{\dagger}$.
      }:
      \begin{equation*}
        \begin{split}
          & \int\limits_{\uptau_i}^{\uptau_f} \dd{\uptau} \eval{\qty(
          \mathcal{T}_{++}( \uptau + \upsigma ) + \mathcal{T}_{--}( \uptau -
          \upsigma ) )}_{\upsigma = 0}^{\upsigma = \uppi}
          \\
          & = - i \frac{T}{4} \int \Updelta \uptau \qty( 2 \eval{\uppsi_{+,\, i}^*
          \lrpartial{\uptau} \uppsi_+^i}^{\upsigma = \uppi}_{\upsigma = 0} -
          \eval{\uppsi_{+,\, i}^* \qty( \R{t}^{\dagger} \lrpartial{\uptau}
          \R{t})^i_j \uppsi_+^j}_{\upsigma = 0} ) \neq 0,
        \end{split}
      \end{equation*}
      while the corresponding condition for the Hamiltonian $\En$:
      \begin{equation*}
        \begin{split}
          & \int\limits_{\uptau_i}^{\uptau_f} \dd{\uptau} \eval{\qty(
          \mathcal{T}_{++}( \uptau + \upsigma ) - \mathcal{T}_{--}( \uptau -
          \upsigma ) )}_{\upsigma = 0}^{\upsigma = \uppi}
          \\
          & = - i \frac{T}{4} \int \Updelta \uptau \qty( \eval{\uppsi_{+,\, i}^*
          \qty( \R{t}^{\dagger} \lrpartial{\uptau} \R{t})^i_j
          \uppsi_+^j}_{\upsigma = 0} ) = 0 \qif \qty( \uptau_i, \uptau_f ) \in
          \qty( \uptauhat_t, \uptauhat_{t-1} ).
        \end{split}
      \end{equation*}
      In both cases we used the shorthand graphical notation
      \begin{equation*}
          \int \Updelta \uptau = \qty( \int\limits_{\uptau_i}^{\uptauhat_{t_0}}
          + \sum\limits_{t = t_0}^{t_N - 1}
          \int\limits_{\uptauhat_t}^{\uptauhat_{t+1}} +
          \int\limits_{\uptauhat_N}^{\uptau_f} ) \dd{\uptau}
      \end{equation*}
      to simplify the written form of the (non-)conservation rules and to stress
      the piecewise nature of the integration domain due to the presence of the
      defects.

      These relations therefore prove that the generator of the
      $\upsigma$-translations \eqref{eq:momentum} is not conserved in time
      because of the boundary conditions, while the time evolution operator
      $\En$ is only piecewise conserved and therefore globally time dependent.

    \subsection{Basis of Solutions and Dual Modes}

    Suppose to have a complete basis of modes $\uppsi_{n,\, \pm}^i$ such that:
    \begin{equation*}
      \begin{cases}
        \uppsi_{n,\, +}^i( \uptau, 0 ) = \qty( \R{t} )^i_j \uppsi_{n,\, -}^j(
        \uptau, 0 ) & \qfor \uptau \in \qty( \uptauhat_t, \uptauhat_{t-1} )
        \\
        \uppsi_{n,\, +}^i( \uptau, \uppi ) = -\uppsi_{n,\, -}^i( \uptau, \uppi )
        & \qfor \uptau \in \mathds{R}
      \end{cases},
    \end{equation*}
    related to a complete basis of the modes of the ``double field''
    $\Uppsi_n^i$ as in \eqref{eq:double-field-Lorentzian}.
    The modes $\uppsi_n$ (and their counterparts $\Uppsi_n$) are a basis of
    solutions of the equations of motion and the boundary conditions for $\uptau
    \in \mathds{R} \setminus \qty{ \uptauhat_t }_{0 \le t \le N}$. The fields
    $\uppsi^i$ (and the fields $\Uppsi^i$) are then a superposition of such
    modes:
    \begin{equation}
      \uppsi^i_{\pm}( \upxi_{\pm} ) = \sum\limits_{n \in \mathds{Z}} b_n
      \uppsi^i_{n,\, \pm}( \upxi_{\pm} ) \quad \Rightarrow \quad \Uppsi^i( \upxi
      ) = \sum\limits_{n \in \mathds{Z}} b_n \Uppsi^i_n( \upxi ).
      \label{eq:usual-mode-expansion}
    \end{equation}

    In order to extract the ``coefficients'' $b_n$ we first introduce the dual
    basis $\dual{\uppsi}_{n,\, \pm}$ (and $\dual{\Uppsi}_n$) in an abstract sense
    such that:
    \begin{itemize}
      \item the dual fields $\dual{\uppsi}_{n,\, \pm}$ (and $\dual{\Uppsi}_n$)
        must be solutions to the equations of motion,
      \item the dual fields $\dual{\uppsi}_{n,\, \pm}$ (and $\dual{\Uppsi}_n$) can
        differ from $\uppsi_{n,\, \pm}$ (and $\Uppsi_n$) in their behavior at the
        boundary,
      \item the functional form of $\dual{\uppsi}_{n,\, \pm}$ (and
        $\dual{\Uppsi}_n$) is fixed by the request of time invariance of the
        usual anti-commutation relations $\qty[ b_n, b_m^{\dagger} ]_+$
        (that is $b_n$ and $b_n^{\dagger}$ can evolve in time, but their
        anti-commutation relations must remain constant).
    \end{itemize}
    We then define the conserved product for the ``double fields''
    \eqref{eq:conserved-product-double-field} in such a way that:
    \begin{equation}
      \eval{\lconsprod{\dual{\Uppsi}_n}{\Uppsi_m}}_{\uptau = \uptau_0} =
      \mathcal{N} \int\limits_0^{2\uppi} \dd{\upsigma}
      \dual{\Uppsi}_{n,\, i}^{*}(\uptau + \upsigma )
      \Uppsi_m^i( \uptau + \upsigma ) = \updelta_{n,m}
      .
      \label{eq:conserved-product-dual-basis}
    \end{equation}
    In the previous expression we changed the notation of the product in order
    to stress that we are dealing with the space of solutions
    whose basis is $\qty{ \Uppsi_n }$ and
    a dual space with basis $\qty{ \dual{\Uppsi}_n }$
    which is not required to span entirely the original space but
    only to be a subset of it in order to be able to compute the
    anti-commutation relations among the annihilation and construction
    operators in a well defined way as in
    \eqref{eq:Mink_can_anticomm_rel_ann_des}.
    
    Given the previous product we can extract the operators as
    \begin{eqnarray*}
      \lconsprod{\dual{\Uppsi}_n}{\Uppsi} & = & b_n,
      \\
      \lconsprod{\dual{\Uppsi}_n^*}{\Uppsi^*} & = & b_n^{\dagger}.
    \end{eqnarray*}
    As a consequence of the canonical anti-commutation relations
    \begin{equation*}
      \qty[ \Uppsi^i\qty( \uptau, \upsigma ), \Uppsi^*_j\qty( \uptau, \upsigma'
      ) ]_+ = \frac{2}{T} \updelta^i_j \updelta( \upsigma - \upsigma' ),
    \end{equation*}
    we have then:
    \begin{equation}
      \eval{\qty[ b_n, b_m^{\dagger} ]_+}_{\uptau = \uptau_0} = \frac{2}{T}
      \mathcal{N} \eval{\lconsprod{\dual{\Uppsi}_n}{\dual{\Uppsi}_m}}_{\uptau =
        \uptau_0}.
      \label{eq:Mink_can_anticomm_rel_ann_des}
    \end{equation}

    As per its definition, the product \eqref{eq:conserved-product-dual-basis}
    is time independent as long as the integrand $\dual{\Uppsi}_n^* \Uppsi_m$ is
    free of singularities at $\uptau = \uptauhat_t$ for $t = 1, 2, \dots, N$.
    Such request on the dual basis automatically fixes its possible form.
    Clearly this does not exclude the possibility to have singularities in
    $\Uppsi_m$ or $\dual{\Uppsi}_n$ separately:
    they are instead deeply
    connected to the boundary changing primary operator hidden
    in the discontinuity of the boundary conditions, that is different
    singularities will be shown to be in correspondence to the excited spin fields.

    Using the definition of the conserved product and defining the fields to
    fulfill some basic requirements we therefore moved the focus from finding a
    consistent definition of the Fock space to the construction of
    the dual basis of modes.
    This task is easier to address in a Euclidean
    formulation and indeed this is the way we will pursue.

  \section{Point-like Defect CFT: the Euclidean Formulation}
  \label{sec:eucl_formulation}

  The main motivation behind the Euclidean reformulation of the previous
  sections is the fact that the solution to the equations of motion on the
  Euclidean strip (or in the complex plane) might be easier to study than its
  Lorentzian worldsheet form. This is specifically the case when $\R{t} \in
  \mathrm{U}(1)^{\Nf} \subset \UN$ on which we shall focus in this paper. The
  presence of a time dependent Hamiltonian is however not completely standard
  and we can neither blindly apply the usual Wick rotation nor the usual CFT
  techniques. We will then be a bit pedantic in order not to miss anything.

  In the following two subsections we
  focus on coordinate changes from the strip to the upper plane not relying on
  the CFT properties since we have not shown that the theory is a CFT. We then
  find the explicit expression of modes which satisfy the equations of motion and the boundary
  conditions and compute the dual modes. Finally we show the algebra of the
  creation and annihilation operators.
  This step is conceptually separated from the definition of the Fock space
  where this algebra is represented: we will in fact take care of it in the
  following sections.

    \subsection{Fields on the Strip}

    Performing the Wick rotation as $\uptau_E = i \uptau$ such that $e^{i
    S_M}=e^{-S_E}$ the Minkowskian action \eqref{eq:cft-action}
    becomes\footnote
    {
      We define the coordinates $\upxi = \uptau_E + i \upsigma$, $\bar \upxi =
      \uptau_E - i \upsigma$ such that $\bupxi = \upxi^*$ and:
      $
        \partial_{\upxi} = \pdv{\upxi} = \frac{1}{2} \qty( \pdv{\uptau_E} -
        i\pdv{\upsigma} ),
      $
      $
        \partial_{\bupxi} = \pdv{\bupxi} = \frac{1}{2} \qty( \pdv{\uptau_E} + i
        \pdv{\upsigma} ).
      $
    }:
    \begin{equation}
      \SE =  \frac{T}{2} \iint \dd{\upxi} \dd{\bupxi}
      \frac{1}{2}~
      \mqty(
      \huppsi_{E,\, +,\, i}^* \lrpartial{\bupxi} \huppsi_{E,\, +}^i
      + \huppsi_{E,\, -,\, i}^* \lrpartial{\upxi} \huppsi_{E,\, -}^i
      )
      ,
      \label{eq:S_Eu_strip}
    \end{equation}
    where the Euclidean fermion on the strip is connected to the Minkowskian
    formulation through
    \begin{equation*}
      \huppsi_{E,\, \pm}^i(\upxi,\bupxi)=
      \uppsi_{\pm}^i(-i\upxi,-i\bupxi).
    \end{equation*}
    As a consequence, the Euclidean ``complex conjugation'' $\star$ (which can
    be defined off-shell) acts as
    \begin{equation}
      \qty[ \huppsi_{E,\, \pm}^i(\upxi, \bupxi) ]^\star = \huppsi_{E,\, \pm
      i}^*(-\bupxi, -\upxi).
      \label{eq:off-shell-Hermitian-conjugate}
    \end{equation}

    The equations of motion are as usual
    \begin{eqnarray*}
      \partial_{\upxi} \huppsi_{E,\, -}^i( \upxi, \bupxi ) = & \partial_{\bupxi}
      \huppsi_{E,\, +}^i( \upxi, \bupxi ) = & 0,
      \\
      \partial_{\upxi} \huppsi_{E,\, -,\, i}^*( \upxi, \bupxi ) = &
      \partial_{\bupxi} \huppsi_{E,\, +,\, i}^*( \upxi, \bupxi ) = & 0,
    \end{eqnarray*}
    whose solutions are the holomorphic functions $\huppsi_{E,\, +}( \upxi )$ and
    $\huppsi_{E,\, -}( \bupxi )$ (and $\huppsi_{E,\, +}^*( \upxi )$ and
    $\huppsi_{E,\, -}^*( \bupxi )$). In these coordinates the boundary conditions
    \eqref{eq:boundary-conditions-solutions} translate to:
    \begin{equation}
      \begin{cases}
        \huppsi_{E,\, -}^i( \uptau_E - i 0^+ ) &= \qty( \R{t} )^i_j \huppsi_{E,\,
        +}^j(\uptau_E + i 0^+ )
        \\
        \huppsi_{E,\, -,\, i}^{*}( \uptau_E - i 0^+ ) &= \qty( \R{t}^* )_i^j
        \huppsi_{E,\, +,\, j}^*(\uptau_E + i 0^+ )
      \end{cases}
      \label{eq:bc_eu_strip}
    \end{equation}
    for $\uptau_E \in \qty( \uptauhat_{E t}, \uptauhat_{E t-1} )$ and
    \begin{equation*}
      \begin{cases}
        \huppsi_{E,\, -}^i( \uptau_E - i \uppi ) &= -\huppsi_{E,\, +}^i( \uptau_E + i
        \uppi )
        \\
        \huppsi_{E,\, -,\, i}^*( \uptau_E - i \uppi ) &= -\huppsi_{E,\, +,\, i}^*( \uptau_E
        + i \uppi )
      \end{cases},
    \end{equation*}
    where $t = 1, 2, \dots, N$ and $\uptauhat_{E,\, t}$ are the
    Wick-rotated locations of the $N$ zero-dimensional defects,
    analytically continued to a real value.

    The conserved product on the strip needs a slight change in the definition
    and becomes:
    \begin{equation}
      \consprod{\widehat{\upalpha}^*_E}{\widehat{\upbeta}_E}
      = \mathcal{N} \int\limits_0^{\uppi}
      \dd{\upsigma}
      \qty( \widehat{\upalpha}^*_{E,\, +,\, i} \widehat{\upbeta}_{E,\, +}^i
      +
      \widehat{\upalpha}^*_{E,\, -,\, i}  \widehat{ \upbeta}_{E,\, -}^i ),
      \label{eq:euclidean-conserved-product-strip}
    \end{equation}
    where $\widehat{\upalpha}^*_E$ and $\widehat{\upbeta}_E$ are the Euclidean
    counterparts of the generic solutions in the original definition of the
    product in \eqref{eq:conserved-product}. In the Euclidean context we have
    to explicitly write $\widehat{\upalpha}^*_E$ because it is no longer
    the ``complex conjugate'' of $\widehat{\upalpha}_E$ in the traditional sense
    but the product is conserved only when it couples two
    solutions which have different boundary conditions as in
    \eqref{eq:bc_eu_strip}.

    The definition of the stress-energy tensor in
    \eqref{eq:stress-energy-tensor-lightcone}
    requires a change in the numerical factor in order to use the usual CFT
    normalization\footnote{
      The canonical coefficient in front of the CFT stress-energy tensor is
      such that the Euclidean Hamiltonian $\Lvir{0}$ is normalized such that
      \begin{equation*}
        \mathcal{T}_{\upzeta\upzeta}( \upzeta ) = \sum_n \Lvir{n}
        e^{-n \upzeta}
      \end{equation*}
      (we have anticipated the double strip notation defined in the next
      subsection for simplicity) then 
      \begin{equation*}
         \En_E = \Lvir{0}
         = \int\limits_{0}^{2\uppi} \frac{\dd{\upphi}}{2 \uppi}
         \mathcal{T}_{\upzeta \upzeta}( \uptau_E + i \upphi )
      \end{equation*}
     therefore $\mathcal{T}_{\upzeta\upzeta}( \upzeta ) = 2 \uppi
     \mathcal{T}^{(can)}_{\upzeta\upzeta}( \upzeta )$.
    }
    and  becomes (introducing a spacetime variable central charge as well):
    \begin{equation*}
      \begin{split}
        \mathcal{T}_{\upxi \upxi}( \upxi ) =
        &- \frac{\uppi T}{2} \huppsi_{E,\, +,\, i}^*( \upxi ) \lrpartial{\upxi}
        \huppsi^i_{E,\, +}( \upxi ) + \widehat{\mathcal{C}}(\upxi ),
        \\
        \mathcal{T}_{\bupxi \bupxi}( \bupxi ) = 
        &- \frac{\uppi T}{2} \huppsi_{E,\, -,\, i}^*( \bupxi ) \lrpartial{\bupxi}
        \huppsi^i_{E,\, -}( \bupxi ) + \widehat{\overline{\mathcal{C}}}( \bupxi ),
      \end{split}
    \end{equation*}
    where $\widehat{{\mathcal{C}}}$ and $\widehat{\overline{\mathcal{C}}}$
    are the leftover terms after the regularization of the singularities due to
    the normal ordering.

    The canonical anti-commutation relations are then
    \begin{equation*}
      \eval{
        \qty[
          \huppsi_{E,\, \pm}^i( \upxi_1, \bupxi_1)
          ,
          \huppsi_{E,\, \pm,\, j}^*( \upxi_2, \bupxi_2 )
        ]_+
      }_{\Re\upxi_1 = \Re\upxi_2}
      =
      \frac{2}{T} \updelta^i_j \updelta\qty( \Im\upxi_1 - \Im\upxi_2 ).
    \end{equation*}

    Given the Euclidean modes $\huppsi^i_{E,\, \pm,\, n}$ and $\huppsi^*_{E,\, \pm,\, n,\, i}$
    (where $n \in \mathds{Z}$) we can then define the dual modes
    $\dual{\huppsi}^i_{E,\, n}$ and $\dual{\huppsi}^*_{E,\, n,\, i}$ such that the
    conserved product \eqref{eq:euclidean-conserved-product-strip} between them
    gives:
    \begin{equation*}
      \lconsprod{\dual{\huppsi}^*_{E,\, n}}{\huppsi_{E,\, m}} =
      \lconsprod{\dual{\huppsi}_{E,\, n}}{\huppsi^*_{E,\, m}} =
      \updelta_{n,m}.
    \end{equation*}

    We can then expand the fields as
    \begin{equation*}
      \begin{cases}
        \huppsi^i_{E,\, +}(\upxi) & = \sum\limits_{n \in \mathds{Z}} b_n
        \huppsi^i_{E,\, +,\, n}(\upxi)
        \\
        \huppsi^i_{E,\, -}(\bupxi) & = \sum\limits_{n \in \mathds{Z}} b_n
        \huppsi^i_{E,\, -,\, n}(\bupxi)
      \end{cases}
    \end{equation*}
    and
    \begin{equation*}
      \begin{cases}
        \huppsi^*_{E,\, +,\, i}(\upxi) & = \sum\limits_{n \in \mathds{Z}} b^*_n
        \huppsi^*_{E,\, +,\, n,\, i}(\upxi)
        \\
        \huppsi^*_{E,\, -,\, i}(\bupxi) & = \sum\limits_{n \in \mathds{Z}} b^*_n
        \huppsi^*_{E,\, -,\, n,\, i}(\bupxi)
      \end{cases}
    \end{equation*}
    in order to extract the operators through the conserved product
    \begin{equation*}
      b_n =   \lconsprod{\dual{\huppsi}^*_{E,\, n}}{\huppsi_{E}},
      \qquad
      b^*_n = \lconsprod{\dual{\huppsi}_{E,\, n}}{\huppsi^*_{E}},
    \end{equation*}
    and get the anti-commutation relations at fixed Euclidean time as
    \begin{equation*}
      \eval{ \qty[ b_n, b^*_m ]_+ }_{\uptau_E = \uptau_{E,\, 0}}
      =
      \frac{2 \mathcal{N}}{T}
      \lconsprod{\dual{\huppsi}^*_{E,\, n}}{\dual{\huppsi}_{E,\, m}}.
    \end{equation*}

    \subsection{Double Strip Formalism and Doubling Trick}

    It is natural to use the usual doubling trick on the strip in order to
    simplify the previous expressions by gluing the holomorphic and
    anti-holomorphic fields along the $\upsigma = \uppi$ boundary.
    Define the coordinate $\upzeta = \uptau_E + i \upphi$ with $0 \le \upphi \le
    2\uppi$, we then have
    \begin{equation*}
      {\hUppsi}( \upzeta ) =
      \begin{cases}
        \huppsi_{E,\, +}( \upzeta) 
        &\qfor \upphi = \upsigma \in \qty[ 0, \uppi ]
        \\
        -\huppsi_{E,\, -}( \upzeta - 2 \uppi i)
        &\qfor \upphi = 2\uppi - \upsigma \in \qty[ \uppi, 2 \uppi ]
      \end{cases}
    \end{equation*}
    on-shell (and similarly for ${\hUppsi}^*( \upzeta )$
    with the substitution ${\huppsi}_{E,\, \pm} \to {\huppsi}_{E,\, \pm}^*$).
    The ``complex conjugation'' $\star$ acts on the off-shell double fields as
    \begin{equation*}
      \qty[ \hUppsi^i(\upzeta,\bupzeta) ]^\star
      = \hUppsi_i^*(-\bupzeta, -\upzeta),
    \end{equation*}
    while the boundary conditions are translated into
    \begin{equation*}
      \begin{cases}
        \hUppsi^i( \uptau_E + 2 \uppi i^- )  =  -\qty( \R{t} )^i_j
        \hUppsi^j( \uptau_E + i 0^+ )
        \\
        \hUppsi^{* i}( \uptau_E + 2 \uppi i^- ) =  -\qty( \R{t}^* )^i_j
        \hUppsi^{* j}( \uptau_E + i 0^+ )
      \end{cases}
    \end{equation*}
    for $\uptau_E \in \qty( \uptauhat_{E,\, t}, \uptauhat_{E,\, t-1} )$.
    The conserved product can then be defined as
    \begin{equation*}
      \consprod{\widehat{A}^*}{\widehat{B}}
      = \mathcal{N}
      \int\limits_{0}^{2\uppi}
      \dd{\upphi} \widehat{A}^{*}_{i}(\uptau_E + i \upphi )
      \widehat{B}^i( \uptau_E + i \upphi ),
    \end{equation*}
    where $\widehat{A}^*$ and $\widehat{B}$ are the double fields connected
    to $\widehat{\upalpha}^*_E$ and $\widehat{\upbeta}_E$ in the previous
    definition on the strip. The holomorphic stress-energy tensor is then
    \begin{equation*}
        \mathcal{T}_{\upzeta \upzeta}( \upzeta ) = - \frac{\uppi T}{2}
        \hUppsi_{i}^*( \upzeta ) \lrpartial{\upzeta} \hUppsi^i( \upzeta ) 
        + \widehat{\mathcal{C}}(\upzeta )
    \end{equation*}
    and the canonical anti-commutation relations are now
    \begin{equation*}
      \eval{
        \qty[
          \hUppsi^i( \upzeta_1 )
          ,
          \hUppsi_{j}^*( \upzeta_2 )
        ]_+
      }_{\Re\upzeta_1 = \Re\upzeta_2}
      =
      \frac{2}{T} \updelta^i_j \updelta\qty( \Im\upzeta_1 - \Im\upzeta_2 ).
    \end{equation*}
    
    The advantage the double fields formulation is in the mode expansion of the
    fields which clarifies that only one coefficient $b_n$ (or $b_n^*$) is
    needed for both $\uppsi_{E,\, +}$ and $\uppsi_{E,\, -}$ (or for both $\uppsi^*_{E,\,
    +}$ and $\uppsi^*_{E,\, -}$).  In fact, given the Euclidean modes
    $\hUppsi^i_{n}$ and $\hUppsi^*_{n,\, i}$ (where $n \in \mathds{Z}$), we can
    define the dual modes $\dual{\hUppsi}^i_{n}$ and $\dual{\hUppsi}^*_{n,\, i}$
    such that
    \begin{equation*}
      \lconsprod{\dual{\hUppsi}^*_{n}}{\hUppsi_{m}} =
      \lconsprod{\dual{\hUppsi}_{n}}{\hUppsi^*_{m}} =
      \updelta_{n,m},
    \end{equation*}
    and expand the double fields as
    \begin{equation*}
      \hUppsi^i(\upzeta) = \sum\limits_{n \in \mathds{Z}} b_n
      \hUppsi^i_{n}(\upzeta),
      \qquad
      \hUppsi^*_{i}(\upzeta) = \sum\limits_{n \in \mathds{Z}} b^*_n
      \hUppsi^*_{n}(\upzeta)
    \end{equation*}
    We then extract the operators as
    \begin{equation}
      b_n =   \lconsprod{\dual{\hUppsi}^*_{n}}{\hUppsi},
      \qquad
      b^*_n = \lconsprod{\dual{\hUppsi}_{n}}{\hUppsi^*}
      \label{eq:upper-half-extraction}
    \end{equation}
    and finally get the anti-commutation relations as
    \begin{equation*}
      \eval{ \qty[ b_n, b^*_m ]_+ }_{\uptau_E = \uptau_{E,\, 0}}
      =
      \frac{2 \mathcal{N}}{T} \lconsprod{\dual{\hUppsi}^*_{n}}{\dual{\hUppsi}_m}.
    \end{equation*}

    \subsection{Fields on the Upper Half Plane}

    \begin{figure}
      \centering
      \includegraphics{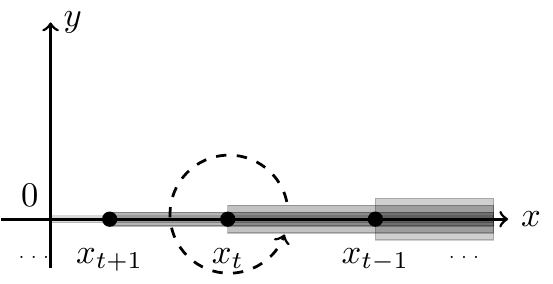}
      \caption{Due to the conformal transformation from the (double) strip to
      the complex plane, fields are glued on the $x < 0$ semi-axis, while there
      are non trivial discontinuities (in the figure they are represented by
      strips with different values of opacity) for $x_t < x < x_{t-1}$ for $t = 1, 2,
      \dots, N$ and where $x_t = \exp( \uptauhat_{E,\, t} )$.}
      \label{fig:complex-plane}
    \end{figure}

    To perform the actual computations we shall however consider another set of
    coordinates on the upper half $\mathcal{H}$ of the complex plane:
    \begin{equation*}
      u = e^{\upxi} \in \mathcal{H} = \qty{ w \in \mathds{C} \mid \Im w \ge 0 },
    \end{equation*}
    where $\upxi = \uptau_E + i \upsigma$ and $\upsigma \in \qty[ 0, \uppi ]$
    define the usual strip, or on the entire complex plane:
    \begin{equation*}
      z = e^{\upzeta} \in \mathds{C},
    \end{equation*}
    where $\upzeta = \uptau_E + i \upphi$ and $\upphi \in \qty[0, 2\uppi ]$
    define the double strip.
    Under this change of coordinates the Euclidean action \eqref{eq:S_Eu_strip}
    becomes
    \begin{equation*}
      \begin{split}
        \SE & = \frac{T}{2} \iint \dd{u}\dd{\bu} \frac{1}{2}
              \qty(
                \frac{1}{u} \huppsi_{E,\, +,\, i}^* \lrpartial{\bu} \huppsi_{E,\, +}^i
                +
                \frac{1}{\bu} \huppsi_{E,\, -,\, i}^* \lrpartial{u} \huppsi_{E,\, -}^i
              )
        \\
            & = \frac{T}{2} \iint \dd{u}\dd{\bu} \frac{1}{2}
              \qty(
                \uppsi_{E,\, +,\, i}^* \lrpartial{\bu} \uppsi_{E,\, +}^i
                +
                \uppsi_{E,\, -,\, i}^* \lrpartial{u} \uppsi_{E,\, -}^i
              ),
      \end{split}
    \end{equation*}
    where we have naturally introduced the off-shell field redefinitions
    \begin{equation}
      \uppsi_{E,\, +}^i(u, \bu) =
      \frac{1}{\sqrt{u}} \huppsi_{E,\, +}^i( \upxi, \bupxi ),
      \qquad
      \uppsi_{E,\, -}^i(u, \bu) =
      \frac{1}{\sqrt{\bu}} \huppsi_{E,\, -}^i( \upxi, \bupxi ).
      \label{eq:euclidean-off-shell-redefinitions}
    \end{equation}
    This way, in the Euclidean context, fields with the hat sign on top represent
    strip and double strip definitions, while fields without the hat sign are
    defined on $\mathcal{H}$ or $\mathds{C}$. We could have anticipated these
    redefinitions from a CFT argument where
    \begin{equation*}
      \uppsi( u ) = \eval{\qty( \dv{u}{\upxi} )^{-\frac{1}{2}}
        {\huppsi}(\upxi)}_{\upxi = \ln( u )},
    \end{equation*}
    but we cannot and do not rely on CFT properties since we have not shown
    that the theory is a CFT yet. Notice that this is the result one would expect
    from the engineering dimension: in this case it works since the theory is
    essentially free. Using the redefinitions
    \eqref{eq:euclidean-off-shell-redefinitions}, the ``complex conjugation''
    $\star$ then becomes
    \begin{equation*}
      \qty[ \uppsi_{E,\, +,\, i}( u, \bu ) ]^\star
      =
      \frac{1}{\bu} \uppsi_{E,\, +,\, i}^*\qty( \frac{1}{\bu}, \frac{1}{u} ),
      \qquad
      \qty[ \uppsi_{E,\, -,\, i}( u, \bu ) ]^\star
      =
      \frac{1}{u} \uppsi_{E,\, -,\, i}^*\qty( \frac{1}{\bu}, \frac{1}{u} ).
    \end{equation*}

    When we choose the cut of the square root on the real negative
    axis the boundary conditions are translated into
    \begin{equation*}
      \begin{cases}
        \uppsi_{E,\, -}^i( x - i 0^+ ) =
        \qty( \R{t} )^i_j \uppsi_{E,\, +}^j( x + i 0^+ )
        \\
        \uppsi^{*}_{E,\, -,\, i}( x - i 0^+ ) =
        \qty( \R{t}^* )_i^j \uppsi^{*}_{E,\, +,\, j}( x + i 0^+ )
      \end{cases}
    \end{equation*}
    for $x \in \qty( x_t, x_{t-1} )$, where $x_t = \exp( \uptauhat_{E,\, t} ) > 0$,
    and
    \begin{equation*}
      \uppsi_{E,\, -}^i( x - i 0^+ ) = \uppsi_{E,\, +}^i( x + i 0^+ ),
      \qquad
      \uppsi^{*}_{E,\, -,\, i}( x - i 0^+ )  =  \uppsi^{*}_{E,\, +,\, i}( x + i 0^+ )
    \end{equation*}
    for $x<0$.

    The product \eqref{eq:euclidean-conserved-product-strip} is then
    \begin{equation}
      \consprod{\upalpha^*}{\upbeta}
      = -i \mathcal{N} \int\limits_{\substack{\abs{u} = \exp( \uptauhat_E ), \\
      0 \le \Im u \le \uppi}}
      \qty[
        \upalpha^*_{+,\, i}(u) \upbeta_+^i(u) \dd{u}
        -
        \upalpha^*_{-,\, i}(\bu) \upbeta_-^i(\bu) \dd{\bu}
      ],
      \label{eq:prod_H}
    \end{equation}
    and the stress-energy tensor\footnote{
      Rewriting the operator part of the stress-energy tensor from the strip
      formulation into the coordinates on $\mathcal{H}$ we actually get
      \begin{equation*}
        \mathcal{T}_{\upxi \upxi}( \upxi(u) )= u^2 \mathcal{T}_{u u}( u ).
      \end{equation*}
      The reason of the presence of $u^2$ can be understood in two ways. Using
      GR we know that $\mathcal{T}_{\upxi \upxi}( \upxi ) (d \upxi)^2
      =\mathcal{T}_{u u}( u ) (d u)^2$. Another more physical way is to
      notice that a translation in $\upxi$ is a dilatation of $u$: the
      infinitesimal generator of $\upxi$ translation must be the infinitesimal
      generator of $u$ dilatation, i.e.
      \begin{equation*}
        P_\upxi\sim \int d\upsigma~\mathcal{T}_{\upxi \upxi} \sim D_u \sim \int
        du u \mathcal{T}_{u u}.
      \end{equation*}
    } becomes:
    \begin{equation*}
      \begin{split}
        \mathcal{T}_{u u}( u ) =
        &- \frac{\uppi T}{2} \uppsi_{E,\, +,\, i}^*( u ) \lrpartial{u}
        \uppsi^i_{E,\, +}( u ) + \widehat{\mathcal{C}}(u ),
        \\
        \mathcal{T}_{\bu \bu}( \bu ) =
        &- \frac{\uppi T}{2} \uppsi_{E,\, -,\, i}^*( \bu ) \lrpartial{\bu}
        \uppsi^i_{E,\, -}( \bu ) + \widehat{\overline{\mathcal{C}}}( \bu ).
      \end{split}
    \end{equation*}
    Finally the anti-commutation relations are
    \begin{equation*}
      \begin{cases}
        \eval{
          \qty[
            \uppsi_{E,\, +}^i( u_1, \bu_1 )
            ,
            \uppsi_{E,\, +,\, j}^*( u_2, \bu_2 )
          ]_+
        }_{\abs{u_1} = \abs{u_2}}
        & =
        \frac{2}{T u_1} \updelta^i_j \updelta( \arg(u_1) - \arg(u_2) )
        \\
        \eval{
          \qty[
            \uppsi_{E,\, -}^i( u_1, \bu_1 )
            ,
            \uppsi_{E,\, -,\, j}^*( u_2, \bu_2 )
          ]_+
        }_{\abs{u_1} = \abs{u_2}}
        & =
        \frac{2}{T \bu_1} \updelta^i_j \updelta( \arg(u_1) - \arg(u_2) ),
      \end{cases}
    \end{equation*}
    which despite the strange look of the expression are perfectly compatible with
    the definition \eqref{eq:upper-half-extraction} leading to:
    \begin{equation*}
      \qty[ b_n, b^*_m ]_+
      =
      \frac{2 \mathcal{N}}{T}
      \lconsprod{\dual{\huppsi}^*_{E,\, n}}{\dual{\huppsi}_{E,\, m}}
      =
      \frac{2 \mathcal{N}}{T}
      \lconsprod{\dual{\uppsi}^*_{E,\, n}}{\dual{\uppsi}_{E,\, m}}
    \end{equation*}
    when the product $\lconsprod{\cdot}{\cdot}$ is defined according to
    \eqref{eq:prod_H}, we expand the fields in modes as
    \begin{equation*}
      \begin{cases}
        \uppsi^i_{E,\, +}(u) =
        \sum\limits_{n \in \mathds{Z}} b_n \uppsi^i_{E,\, +,\, n}(u)
        \\
        \uppsi^i_{E,\, -}(\bu) =
        \sum\limits_{n \in \mathds{Z}} b_n \uppsi^i_{E,\, -,\, n}(\bu)
      \end{cases}
    \end{equation*}
    and
    \begin{equation*}
      \begin{cases}
        \uppsi^*_{E,\, +,\, i}(u) =
        \sum\limits_{n \in \mathds{Z}} b^*_n \uppsi^*_{E,\, +,\, n,\, i}(u)
        \\
        \uppsi^*_{E,\, -,\, i}(\bu) =
        \sum\limits_{n \in \mathds{Z}} b^*_n \uppsi^*_{E,\, -,\, n,\, i}(\bu)
      \end{cases}
    \end{equation*}
    and $\dual{\uppsi}_{E,\, n}$ and $\dual{\uppsi}^*_{E,\, n}$ are the corresponding
    dual modes on the upper half plane.   

    \subsection{Fields on the Complex Plane and the Doubling Trick}
    
    As in the double strip formulation, we can use the doubling trick in order
    to define the fields on the subset $\mathds{C} \setminus \qty[ x_N, x_1 ]$:
    \begin{equation*}
      \Uppsi( z ) =
      \begin{cases}
        \uppsi_{E,\, +}( u)   &\qfor z = u \in \mathcal{H} \setminus \qty[ x_N, x_1 ]
        \\
        \uppsi_{E,\, -}( \bu) &\qfor z = \bu \in \mathcal{H}^* \setminus \qty[ x_N, x_1 ]
      \end{cases}
    \end{equation*}
    where $z = \exp( \uptau_E + i \upphi ) = x + i y$ and
    $
      \mathcal{H}^* = \qty{ w \in \mathds{C} \mid \Im w \le 0 }
    $
    (the same goes for $\Uppsi^*$ with the exchange $\uppsi_{E,\, \pm} \to
    \uppsi_{E,\, \pm}^*$).

    In this case the ``complex conjugation'' $\star$ acts off-shell as
    \begin{equation}
      \qty[ \Uppsi^i( z, \bz ) ]^\star = \frac{1}{\bar
      z}\Uppsi_i^*\qty(\frac{1}{\bz}, \frac{1}{z})
      \label{eq:complex-plane-conjugate}
    \end{equation}
    and the boundary conditions are
    \begin{equation}
      \begin{cases}
        \Uppsi^i( x - i 0^+ ) & = \qty( \R{t} )^i_j \Uppsi^j( x + i 0^+ ),
        \\
        \Uppsi^{*\, i}( x - i 0^+ ) & = \qty( \R{t}^* )^i_j \Uppsi^{*\, j}( x
        + i 0^+ ),
      \end{cases}
      \label{eq:boundary-condition-euclidean}
    \end{equation}
    for $x \in \qty( x_t, x_{t-1} )$, where $x_t = \exp( \uptauhat_{E t} ) > 0$
    for $t \in \qty{ 1, 2, \dots, N }$. When $x < 0$ we get
    \begin{equation}
      \begin{cases}
      \Uppsi( x - i 0^+ ) &= \Uppsi( x + i 0^+ ),
      \\
      \Uppsi^*( x - i 0^+ ) &=      \Uppsi^*( x + i 0^+ )
      \end{cases}
      \label{eq:gluing-conditions-euclidean}
    \end{equation}
    instead.

    Given the relations $\dd{z} = i z \dd{\upphi}$
    we can write the conserved product \eqref{eq:prod_H} as:
    \begin{equation}
      \consprod{A^*}{B}
      =
      2\uppi \mathcal{N}
      \oint\limits_{\abs{z} =\exp( \uptau_E )}
      \frac{\dd{z}}{2 \uppi i} A^*_i( z ) B^i( z ),
      \label{eq:conserved-product-complex-plane}
    \end{equation}
    where we explicitly stressed that the integral has to be performed
    at a fixed Euclidean time $\uptau_E$: in the new coordinate on the plane,
    the conserved product becomes a contour integral at a fixed radius from the
    origin.

    In the same way we can recast the stress-energy tensor components
    \eqref{eq:stress-energy-tensor-lightcone} in the new coordinates:
    \begin{equation*}
      \mathcal{T}( z ) = - \frac{\uppi T}{2} \Uppsi^*_i( z ) \lrpartial{z}
      \Uppsi^i( z ) + \mathcal{C}( z ),
    \end{equation*}
    where $\mathcal{T} = \mathcal{T}_{zz}$ for simplicity.

    Finally the canonical anti-commutation relations between the fields
    are:
    \begin{equation*}
      \eval{
        \qty[
          \Uppsi^i( z_1 )
          ,
          \Uppsi_{j}^*( z_2 )
        ]_+
      }_{\abs{z_1} = \abs{z_2}}
      =
      \frac{2}{T z_1} \updelta^i_j \updelta( \arg(z_1) - \arg(z_2) ),
    \end{equation*}
    the fields expansion in modes reads
    \begin{equation}
      \Uppsi^i(z) =
      \sum\limits_{n \in \mathds{Z}} b_n \Uppsi^i_{n}(z),
      \qquad
      \Uppsi^*_{i}(z) =
      \sum\limits_{n \in \mathds{Z}} b^*_n \Uppsi^*_{n. i}(z),
      \label{eq:complex-plane-mode-expansion}
    \end{equation}
    and the anti-commutation relations among the operators are
    \begin{equation*}
      \qty[ b_n, b^*_m ]_+ =
      \frac{2 \mathcal{N}}{T} \lconsprod{\dual{\Uppsi}^*_{n}}{\dual{\Uppsi}_{m}},
    \end{equation*}
    when we introduce the dual modes
    $\dual{\Uppsi}_{n}(z)$ and $\dual{\Uppsi}^*_{n}(z)$ whose normalization is
    \begin{equation*}
      \lconsprod{\dual{\Uppsi}^*_{n}}{{\Uppsi}_{m}}
      =
      \lconsprod{\dual{\Uppsi}_{n}}{{\Uppsi}^*_{m}}
      =
      \updelta_{m,n}.
    \end{equation*}
    
  \section{Algebra of Creation and Annihilation Operators}
  \label{sec:modes_and_algebra}
  
  In this section we find the explicit expression of the modes which
  satisfy the equations of motion and the boundary conditions. We then compute
  the dual fields and finally the algebra of the creators and annihilators.

    \subsection{NS Complex Fermions}
    \label{sec:ns-complex-fermions}

    In order to check that this formalism agrees with known results we start
    from the simplest case at hand: NS complex fermions. Consider the usual
    definition:
    \begin{equation*}
      \begin{cases}
        \uppsi_-^i( \uptau, 0 )
        & =
        \uppsi_+^i( \uptau, 0 ),
        \\
        \uppsi_-^i( \uptau, \uppi )
        & =
        -\uppsi_+^i( \uptau, \uppi )
      \end{cases}
    \end{equation*}
    for $\uptau \in \mathds{R}$, which can be recovered from
    $\eqref{eq:boundary-conditions-solutions}$ setting $\R{t} \equiv \mathds{1}$
    .
    In the Euclidean formulation, we use
    \eqref{eq:boundary-condition-euclidean} and
    \eqref{eq:gluing-conditions-euclidean} to get:
    \begin{equation*}
      \begin{cases}
        \Uppsi( x - i 0^+ ) & =  \Uppsi( x + i 0^+ )
        \\
        \Uppsi^*( x - i 0^+ ) & =  \Uppsi^*( x + i 0^+ )
      \end{cases}
    \end{equation*}
    for $x \in \mathds{R}$.

    In order to recover the definition of the dual modes
    \eqref{eq:conserved-product-dual-basis} using the Euclidean conserved
    product \eqref{eq:conserved-product-complex-plane}, we define:
    \begin{eqnarray*}
      \Uppsi^i_{( n, i_0 )}( z ) & = & \normfac \updelta^i_{i_0}
      z^{-n},
      \\
      \dual{\Uppsi}_{( m, j_0 ),\, j}( z ) & = & \qty( 2 \uppi \mathcal{N}
      \normfac )^{-1} \updelta_{j, j_0} z^{m-1}
                                              ,
    \end{eqnarray*}
    and similarly for $\Uppsi^*$,
    in such a way that
    \begin{equation*}
      \lconsprod{\dual{\Uppsi}_{( n, i_0 )}^*}{\Uppsi_{( m, j_0 )}} =
      \lconsprod{\dual{\Uppsi}_{( m, j_0 )}}{\Uppsi^*_{( n, i_0 )}} =
      \updelta_{n, m} \updelta_{i_0, j_0}
      .
    \end{equation*}

    As a consequence we find
    \begin{equation*}
      \lconsprod{\dual{\Uppsi}_{( n, i_0 )}^*}{\dual{\Uppsi}_{( m, i_1 )}} =
      \frac{1}{2 \uppi \mathcal{N} \mathcal{N}^2_{\Uppsi}}
      \updelta_{i_0, i_1} \updelta_{n + m, 1}.
    \end{equation*}

    Consider the  NS expansion in modes of the double fields:
    \begin{eqnarray*}
      \Uppsi^i( z ) & = & \sum\limits_{n \in \mathds{Z}} \sum\limits_{i_0}
      b_{(n, i_0)} \Uppsi^i_{( n, i_0 )}( z ),
      \\
      \Uppsi^{*}_{ i}( z ) & = & \sum\limits_{n \in \mathds{Z}}
      \sum\limits_{i_0} b^*_{(n, i_0)} \Uppsi^{*}_{( n, i_0 ),\, i}( z ),
    \end{eqnarray*}
    then
    \begin{eqnarray*}
      b_{( n, i_0 )} & = &
                           \lconsprod{\dual{\Uppsi}_{( n, i_0
      )}^*}{\Uppsi},
      \\
      b^*_{( n, i_0 )}
                     & = &
                           \lconsprod
                           {\dual{\Uppsi}_{( n, i_0)}}
                           {\Uppsi^*}
                           ,
    \end{eqnarray*}
    and
    \begin{equation}
      \qty[ b_{( n, i_0 )}, b^*_{( m, j_0 )} ]_+ = \frac{1}{\uppi T
      \normfac^2} \updelta_{i_0, j_0} \updelta_{n + m, 1}.
      \label{eq:ns-algebra}
    \end{equation}

    \subsection{Twisted Complex Fermions: preliminaries}

    We can now move to a more general discussion of $\Nf = 1$ complex fermions
    in the presence of $N$ point-like defects which we will show to be
    primary boundary changing operators (i.e. plain and excited spin fields).
    Let
    \begin{equation*}
      \begin{cases}
        \R{t} & = e^{i \uppi \upalpha_{( t )}} \in \mathrm{U}( 1 )
        \\
        \R{t}^* & = e^{-i \uppi \upalpha_{( t )}} \in \mathrm{U}( 1 )
      \end{cases}
    \end{equation*}
    such that $0 < \upalpha_{( t )} < 2$. We have the boundary conditions:
    \begin{equation*}
      \begin{cases}
        \Uppsi( x - i 0^+ ) & =  e^{i \uppi \upalpha_{( t )}} \Uppsi( x + i 0^+
        )
        \\
        \Uppsi^*( x - i 0^+ ) & = e^{-i \uppi \upalpha_{( t )}} \Uppsi^*( x + i
        0^+ )
      \end{cases},
    \end{equation*}
    for $x \in ( x_t, x_{t-1} )$, and
    \begin{equation*}
      \begin{cases}
        \Uppsi( x - i 0^+ ) & =  \Uppsi( x + i 0^+)
        \\
        \Uppsi^*( x - i 0^+ ) & = \Uppsi^*( x + i 0^+ )
      \end{cases},
    \end{equation*}
    for $x < 0$.
    Also in this case we can refer to
    Figure~\ref{fig:complex-plane} to keep in mind the intuitive picture. The
    boundary conditions can be recast in the form of monodromy
    factors.
    Performing a loop around $x_t$ we find
    \begin{equation*}
      \Uppsi( x_t + \updelta e^{i 0^+} ) = e^{i \uppi \qty( \upalpha_{( t )} -
      \upalpha_{( t + 1 )} )} \Uppsi( x_t + \updelta e^{ 2 \uppi i} ),
    \end{equation*}
    where $\updelta \in \mathds{R}^+$ is small enough\footnote{
      Technically, $0 < \updelta < \min\qty( \abs{x_{t-1} - x_t}, \abs{x_t -
      x_{t+1}} )$.
    }
    and the $\pm$ in the phase represents the
    position relative to the real axis ($+$ is in the upper half plane, while
    $-$ in the lower half plane). Let us define\footnote{
      \label{foot:other_range}
      Notice that the choice of the range for $\eps{t}$ is not unique.
      We can choose $0< \upalpha_{( t )} < 2$ leading to
            $
            \eps{t} = \upalpha_{( t+1 )} - \upalpha_{( t )} + 2 \uptheta( \upalpha_{(
      t )} - \upalpha_{( t+1 )} )
$
      Then in this case $\eps{t} = 2-\beps{t}$ and $\eps{t}, \, \beps{t} \in \qty(
      0, 2 )$.
      We will however stick to the first definition in the following
      sections since it allows to consider the NS case as special. 
    }
    \begin{equation*}
      \eps{t} = \upalpha_{( t+1 )} - \upalpha_{( t )}
      + \uptheta( \upalpha_{(
        t)} -\upalpha_{( t+1 )} - 1 )
      - \uptheta( \upalpha_{( t+1 )} - \upalpha_{(
        t )} - 1 )
    \end{equation*}
    such that
    \begin{equation*}
      -1 < \eps{t} < 1 \qquad \forall t = 1, 2, \dots, N,
    \end{equation*}
    then the previous loop around $x_t$ induces  a monodromy 
    \begin{equation}
            \begin{cases}
        \Uppsi( x_t + \updelta e^{i 0^+} ) & = e^{-i \uppi \eps{t}} \Uppsi( x_t
        + \updelta e^{2 i \uppi^+} )
        \\
        \Uppsi^*( x_t + \updelta e^{i 0^+} ) & = e^{-i \uppi \beps{t}} \Uppsi^*(
        x_t + \updelta e^{2 i \uppi^+} ),
      \end{cases}
      \label{eq:monodromy-factors}
    \end{equation}
    where $\beps{t} =  - \eps{t} \Rightarrow -1 < \beps{t} < 1$
    thus showing a symmetry under the exchange of:
    \begin{equation*}
      \Uppsi \longleftrightarrow \Uppsi^* \qquad \Rightarrow \qquad \eps{t}
      \longleftrightarrow \beps{t}.
    \end{equation*}

      \subsubsection{Usual Twisted Fermions}
      \label{sec:usual-twisted-fermion}

      As it is useful in the discussion of the meaning of the defects,
      we consider the case of one complex fermion in the presence of one twisted
      boundary condition with the defects located at zero and infinity.
      We take $N = 2$ and $x_1 = \infty$ and $x_2 = 0$.
      For simplicity we denote $\epss$ the argument of the monodromy factor
      arising from the presence of the cut on the interval $( 0,
      +\infty)$.

      In
      order to fulfill the requests \eqref{eq:monodromy-factors} we can write the
      modes as:
      \begin{equation}
        \begin{split}
          \Uppsi_n^{( \Es )} & = \normfac z^{-n + \Es},
          \\
          \Uppsi_n^{*\, ( \bEs )} & = \normfac z^{-n + \bEs},
        \end{split}
        \label{eq:usual-twisted-modes}
      \end{equation}
      such that
      \begin{equation*}
        \begin{split}
          \Es = n_{\Es} + \frac{\epss}{2},&\quad n_{\Es} \in \mathds{Z},
          \\
          \bEs = n_{\bEs}  +\frac{\bepss}{2},&\quad n_{\bEs} \in \mathds{Z}.
        \end{split}
      \end{equation*}
      Together with the integer factor $n_{\Es}$ and $n_{\bEs}$ we also define a
      third integer for later convenience\footnote{
        The choice discussed in footnote~\ref{foot:other_range}
        gives  $\LLs = n_{\Es} + n_{\bEs} +1$.   
        We can easily exchange the definitions using
        $\beps{t}^{2nd}=\beps{t}^{1st}+2$ and
        $n^{2nd}_{\bEs}=n^{1st}_{\bEs}-1$.
    }:
      \begin{equation*}
        \LLs = \Es + \bEs = n_{\Es} + n_{\bEs} \in \mathds{Z}.
      \end{equation*}
      In order to extract the creators and annihilators from the conserved product
      \eqref{eq:conserved-product-complex-plane}, we define the dual basis as:
      \begin{eqnarray*}
        \dual{\Uppsi}_n^{( \bEs )}( z ) & = & \frac{1}{2 \uppi \mathcal{N}
        \normfac} z^{n - 1 - \bEs},
        \\
        \dual{\Uppsi}_n^{*\, ( \Es )}( z ) & = & \frac{1}{2 \uppi \mathcal{N}
        \normfac} z^{n - 1 - \Es}.
      \end{eqnarray*}
      This way we compute the usual anti-commutation relations as
      \begin{equation}
        \lconsprod{\dual{\Uppsi}_n^{*\, ( \Es )}}{\dual{\Uppsi}_m^{( \bEs )}} =
        \frac{\updelta_{n + m, 1 + \LLs}}{2 \uppi \mathcal{N} \normfac^2}
        \quad \Rightarrow \quad
        \qty[ b_n, b_m^* ]_+ = \frac{1}{\uppi T \normfac^2} \updelta_{n + m, 1 +
        \LLs},
        \label{eq:twisted-fermion-algebra}
      \end{equation}
      which are constant in time independently of  $\Es$ or $\bEs$
      since the only possible singularities are at $z = 0$ and $z = \infty$.
      We can then expand the fields
      $\Uppsi(z)$ and $\Uppsi^*( z )$ using this basis or the more
      conventional one
      as
      \begin{eqnarray}
        \Uppsi( z )
        & = &
              \sum\limits_{n \in \mathds{Z}} \obE_n \Uppsi_n^{(
              \Es )}( z )
              = \sum\limits_{n \in \mathds{Z}} b_{n
              + n_{\Es}} \Uppsi_n^{( \frac{\epss}{2} )}( z ),
        \label{eq:usual-twisted-mode-expansion}
        \\
        \Uppsi^*( z )
        & = &
              \sum\limits_{n \in \mathds{Z}} \obbE_n
              \Uppsi_n^{*\, ( \bEs )}( z )
              = \sum\limits_{n \in \mathds{Z}} b^{*}_{n + n_{\bEs} } \Uppsi_n^{*\, ( -\frac{\epss}{2}
              )}( z )
              ,
        \label{eq:usual-twisted-mode-expansion-conjugate}
      \end{eqnarray}
      where we have used the shorter notation $b=b^{( \frac{\epss}{2} )}$
      and $b^*=b^{*\,( \frac{\epss}{2} )}$.


      \subsubsection{Generic Case With Defects}
      \label{sec:generic-twisted}

      We now consider one complex fermion in the presence of $N$ defects
      such that the modes satisfy:
      \begin{equation*}
        \Uppsi_n( x_t + \updelta e^{2 \uppi i^+} ) = e^{i \uppi \eps{t}} \Uppsi_n(
        x_t + \updelta e^{i 0^+} )
      \end{equation*}
      for $t = 1, 2, \dots, N$ and $\updelta > 0$.
      We define the basis of    solutions as:
      \begin{eqnarray}
        \Uppsi_n( z ; \{x_t, \E t\} ) & = & \normfac z^{-n} \prodt \qty( 1 - \frac{z}{x_t}
        )^{\E{t}},
        \label{eq:generic-case-basis}
        \\
        \Uppsi^*_n( z ; \{x_t, \bE t\} ) & = & \normfac z^{-n} \prodt \qty( 1 - \frac{z}{x_t}
        )^{\bE{t}},
        \label{eq:generic-case-basis-conjugate}
      \end{eqnarray}
      where we generalise the definition of
      \begin{eqnarray*}
        \E{t} = & n_{\E{t}} + \frac{\eps{t}}{2},&\quad n_{\E{t}} \in \mathds{Z},
        \\
        \bE{t} = & n_{\bE{t}} + \frac{\bepss_{(t)} }{2}&\quad n_{\bE{t}} \in
        \mathds{Z}
      \end{eqnarray*}
      and we define $N$  integer factors:
      \begin{equation*}
        \LL{t} = \E{t} + \bE{t} = n_{\E{t}} + n_{\bE{t}}  \in \mathds{Z}
        ,
      \end{equation*}
      for $t = 1, 2, \dots, N$,
      in analogy to \eqref{eq:conserved-product-complex-plane}.
      From the definition of the conserved product
      \eqref{eq:conserved-product-complex-plane}, we compute the dual basis:
      \begin{eqnarray*}
        \dual{\Uppsi}_n( z ) & = & \frac{1}{2 \uppi \mathcal{N} \normfac}
        z^{n-1} \prodt \qty( 1 - \frac{z}{x_t} )^{-\bE{t}},
        \\
        \dual{\Uppsi}^*_n( z ) & = & \frac{1}{2 \uppi \mathcal{N} \normfac}
        z^{n-1} \prodt \qty( 1 - \frac{z}{x_t} )^{-\E{t}},
      \end{eqnarray*}
      and the conserved products between dual modes:
      \begin{equation*}
          \lconsprod{\dual{\Uppsi}_n^*}{\dual{\Uppsi}_m} = \frac{1}{2 \uppi
          \mathcal{N} \normfac^2} \oint \frac{\dd{z}}{2 \uppi i} z^{n+m-2}
          \prodt \qty( 1 - \frac{z}{x_t} )^{-\LL{t}}
          .
        \end{equation*}
      Notice that the products are radially invariant only if
      \begin{equation}
        \LL{t} \le 0 \qquad \forall t \in \qty{ 1, 2, \dots, N },
        \label{eq:generic-case-negativity-condition}
      \end{equation}
      since the integrand must not present time dependent singularities on the
      integration path, thus
      \begin{equation*}
        \begin{split}
          \lconsprod{\dual{\Uppsi}_n^*}{\dual{\Uppsi}_m} 
          & = \frac{1}{2 \uppi \mathcal{N} \normfac^2} \oint \frac{\dd{z}}{2
          \uppi i} \prodt \sum\limits_{k_t = 0}^{\abs{\LL{t}}}
          \binom{\abs{\LL{t}}}{k_t} \qty( - \frac{1}{x_t})^{k_t} z^{k_t+n+m-2}
          \\
          & = \frac{1}{2 \uppi \mathcal{N} \normfac^2} p_{1-n-m},
        \end{split}
      \end{equation*}
      where we defined
      \begin{equation}
        p_k = \prodt \sum\limits_{k_t = 0}^{\abs{\LL{t}}} \binom{\abs{\LL{t}}}{k_t}
        \qty( - \frac{1}{x_t} )^{k_t} \updelta_{\sumt
        k_t, k}
        \label{eq:generic-conserved-product-factor}
      \end{equation}
      such that
      \begin{eqnarray*}
        p_{0 \le k \le \sumt \abs{\LL{t}}} & \neq & 0,
        \\
        p_{k \le -1} = p_{k \ge \sumt \abs{\LL{t}} + 1} & = & 0.
      \end{eqnarray*}
      We can finally write
      \begin{equation}
        \qty[ b_n, b^*_m ]_+ = \frac{1}{\uppi T \normfac^2} p_{1-n-m},
        \qquad
        1 - \sumt \abs{\LL{t}} \le n + m \le 1.
        \label{eq:generic-case-anti-commutation}
      \end{equation}

  \section{Representation of the Algebra: Definition of the In-Vacuum}
  \label{sec:invacuum}
  
  In the previous section we computed the algebra of the operators
  for different theories. We now define in-vacua where they are
  represented. This is the first step to define the building blocks for
  computing correlation functions. We will show how to recover the usual NS
  vacuum and the usual twisted vacuum with a slightly different twist from the
  usual. Finally we will discuss the vacuum in the presence of a generic number of
  defects.

  In the previous section we have seen that given the monodromies of the defects
  we can have many different singularities.  Since we want to identify the
  defects as (excited) spin fields we want to understand what is the local
  singularity associated with excited twisted vacua. 
  
    \subsection{NS Fermions}

    The case of NS fermions is trivial since there are no defects.
    The in-vacuum can be correctly obtained either by requiring
    $\Uppsi( z )$ and $\Uppsi^*( z )$ to be non singular as $z \to
    0$ when applied on the vacuum
    or by the same request on 
    $\hat{\Uppsi}( \upxi )$ and
      $\hat{\Uppsi}^*( \upxi )$.
In both cases we get the same vacuum which 
    turns out to be $\SL$ invariant:
    \begin{equation}
      b_{( n, i_0 )} \ket{0}_{\SL} = b^*_{( n, i_0 )} \ket{0}_{\SL} = 0,\quad n
      \ge 1.
      \label{eq:NS_SL2_vacuum}
    \end{equation}
    The spectrum of the theory can be constructed acting with operators $b_{( n,
    i_0 )}$ and $b^*_{( n, i_0 )}$ with $n \le 0$.
    
    \subsection{Twisted Fermion}

    Consider the case of the usual twisted fermion in
    Section~\ref{sec:usual-twisted-fermion}. We start from the
    definition of the excited vacuum and work out the way to the
    minimum energy configuration. We will then discuss the result.

      \subsubsection{Excited vacuum}
      \label{sec:twisted-excited-vacuum}

      Define the excited vacuum $\excvacket$ as:
      \begin{equation}
        \obE_n \excvacket = \obbE_n \excvacket = 0,\quad n \ge 1.
        \label{eq:usual-excited-vacuum}
      \end{equation}
      The reason for the introduction of $E$ and $\bar E$ is to be able to
      define this vacuum as above, i.e. with a $n$ range independent
      on  them and, at the same time, 
      to have a non trivial singularity as $z\rightarrow0$ which does
      depend on them, explicitly
      \begin{equation}
        \Uppsi(z) \excvacket \sim z^{\Es} (\dots)
        ,\qquad
        \Uppsi^*(z) \excvacket \sim z^{\bEs} (\dots)
        \label{eq:asymp_beha_Psi_on_exc_vac}
        .
      \end{equation}
      By comparison with \eqref{eq:generic-case-basis} and
      \eqref{eq:generic-case-basis-conjugate}
      this behavior suggests that in the point $x_t$ there is a hidden
      operator which creates $\excvacket$ with $\Es=\E t$ and $\bEs=
      \bE t$.
      
\begin{figure}
        \centering
        \includegraphics{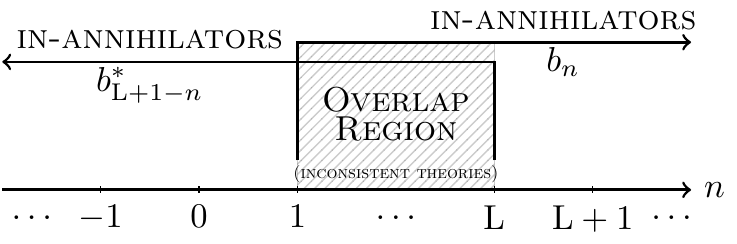}
        \caption{As a consistency condition, we have to exclude the values of
        $\LLs$ for which both $\obE_n$ and $\obbE_{\LLs + 1 - n}$ are in-annihilators
        with a non vanishing anti-commutation relation.}
        \label{fig:inconsistent-theories}
      \end{figure}

      These relations are subject to consistency conditions since
      \begin{equation*}
        \excvacket = \uppi T \normfac^2 \qty[ \obE_n, \obbE_{\LLs + 1 - n} ]_+
        \excvacket,
      \end{equation*}
      that is we cannot have two in-annihilators (namely both $\obE_n$
      and
      $\obbE_{\LLs+ 1 - n}$) with non vanishing anti-commutation relations.
      Specifically we have that (see Figure~\ref{fig:inconsistent-theories} for
      a graphic description):
      \begin{equation*}
        1 \le n \le \LLs
        \quad \Rightarrow \quad
        \obE_n \excvacket = 0,\qquad \obbE_{\LLs + 1 - n} \excvacket = 0,
      \end{equation*}
      that is
      \begin{equation*}
        \excvacket = \uppi T \normfac^2 \qty[ \obE_n, \obbE_{\LLs + 1 - n} ]_+
        \excvacket = 0,
      \end{equation*}
      which is not consistent: the theory does not exist. We shall therefore
      consider only cases such that
      \begin{equation*}
        \LLs \le 0,
      \end{equation*}
      analogously to \eqref{eq:generic-case-negativity-condition}.

      Moreover notice that for $\LLs \le -1$ both $\obE_{\LLs \le n \le 0}$ and
      $\obbE_{\LLs \le n \le 0}$ are in- and out-creation operators:
      in the next section we will show that this case is not acceptable.

      \subsubsection{Minimum Energy Vacuum}
      \label{sec:minimum-energy-vacuum}

      The vacuum $\excvacket$ defined in the previous section is not however
      associated to the lowest energy.
      In fact the usual way to build the vacuum would be
      to require $\Uppsi( z )$ and $\Uppsi^*( z )$ to be non singular as $z \to
      0$ for the in-vacuum so that
$                        \obE_n \twsvacket=0  \qfor n > \Es,
$ and
$                         \obbE_n \twsvacket=0  \qfor n >
        \bEs.
$
      However this procedure almost always fails to give a good  definition of
      the vacuum (it works only for NS fermions).
      For example when $\epss > 0$ we have:
      \begin{equation*}
        0 = \uppi T \normfac^2 \qty[ \obE_{1 + n_{\Es} }, \obbE_{n_{\bEs} }
        ]_+ \twsvacket = \twsvacket,
      \end{equation*}
      which is not consistent since both  $\obE_{1 + n_{\Es} }$
      and $\obbE_{n_{\bEs} }$ are annihilators
      as $1 + n_{\Es}>E$ and $n_{\bEs}>\bar E$.

      The minimum energy vacuum is instead defined in a proper way on
      the strip.
      Requiring that the action of $\hat{\Uppsi}( \upxi )$ and
      $\hat{\Uppsi}^*( \upxi )$
      for $\upxi \to -\infty$ on the vacuum is well defined we get
      \begin{eqnarray*}
        \obE_n \twsvacket & = & 0,\quad n > \Es + \frac{1}{2},
        \\
        \obbE_m \twsvacket & = & 0,\quad m > \bEs + \frac{1}{2}.
      \end{eqnarray*}
      This is a good definition of the vacuum since      
        $-\frac{1}{2} < \frac{\epss}{2}=-\frac{\bepss}{2} < \frac{1}{2}$
        implies that $\obE_n$ and $\obbE_m$ are annihilation operators for
        $n \ge n_{\Es}+1 >\Es + \frac{1}{2}$ and
        $m \ge n_{\bEs}+1 > \bEs + \frac{1}{2}$
        so that
          \begin{equation*}
            0 =
            \uppi T \normfac^2 \qty[ \obE_{n},
            \obbE_{m} ]_+ \twsvacket
            = \updelta_{n + m, \Es+\bEs+1 } \twsvacket = 0
          \end{equation*}
          
          This way we get a consistent definition of the twisted
          vacuum\footnote{
      Notice that the second choice of $\epss$ interval discussed
      in footnote~\ref{foot:other_range}
      needs to distinguish between two cases: $0 < \frac{\epss}{2} <
      \frac{1}{2}$ (and $\frac{1}{2} < \frac{\bepss}{2} < 1$) and $\frac{1}{2} <
      \frac{\epss}{2} < 1$ (and $0 < \frac{\bepss}{2} < \frac{1}{2}$).
    }
    which however is not in general $\SL$ invariant as we will show after
          the construction of the stress-energy tensor.

      \subsubsection{Relation Between Vacua}

      The two vacua $\excvacket$ and $\twsvacket$ are related.
      Consider for example, the case $n_{\Es} \ge 1$ and the definition of the
      vacua:
      \begin{eqnarray*}
        \obE_n \excvacket & = & 0,\quad n \ge 1,
        \\
        \obE_n \twsvacket & = & 0,\quad n \ge 1 + n_{\Es}.
      \end{eqnarray*}
      Then for $1 \le n \le n_{\Es}$ the modes $\obE_n$ act as a annihilation
      operator on $\excvacket$ and as a creation operator on $\twsvacket$:
      \begin{equation}
        \excvacket \propto \obE_{n_{\Es}} \obE_{n_{\Es} - 1} \dots \obE_1 \twsvacket.
        \label{eq:usual-twisted-vacuum-relation}
      \end{equation}
      Moreover, since $\LLs = n_{\Es} + n_{\bEs}  \le 0
      \Rightarrow n_{\bEs} \le -1$, we have:
      \begin{eqnarray*}
        \obbE_m \excvacket & = & 0,\quad m \ge 1,
        \\
        \obbE_n \twsvacket & = & 0,\quad m \ge 1 - \abs{n_{\bEs}},
      \end{eqnarray*}
      which leads for the same argument to:
      \begin{equation}
        \twsvacket \propto \obbE_0 \obbE_1 \dots \obbE_{1 - \abs{n_{\bEs}}}
        \excvacket.
        \label{eq:usual-twisted-fermion-conformal-twisted}
      \end{equation}
      In order to check the consistency of the definition, we require that:
      \begin{equation*}
        \excvacket = \qty( \uppi T \normfac^2 )^{n_{\Es}} \obE_{n_{\Es}}
        \obE_{n_{\Es}-1} \dots \obE_1 \obbE_0 \obbE_1 \dots \obbE_{1 - \abs{n_{\bEs}}}
        \excvacket,
      \end{equation*}
      where the number of $b$ operators has to match the number of $b^*$
      operators:
      \begin{equation}
        n_{\Es} + n_{\bEs}  = \Es + \bEs = \LLs = 0.
        \label{eq:twisted-fermion-consistency}
      \end{equation}
      The same procedure applies also in the case $n_{\Es} \le 0$, leading to
      the same result.

      As a consequence of \eqref{eq:twisted-fermion-consistency}, we can 
      express the twisted vacuum as:
      \begin{eqnarray*}
        \obE_n \twsvacket & = & 0,\quad n \ge 1 + n_{\Es},
        \\
        \obbE_m \twsvacket & = & 0,\quad m \ge 1 - n_{\Es}
.
      \end{eqnarray*}

    \subsection{Generic Case with Defects}
    
    Since the fields in presence of defects behave as NS fields in the
    limit $z \to 0$, we can define the vacuum in the usual fashion
    by requiring a finite limit 
    $
        \lim\limits_{z \to 0} \Uppsi( z ) \Gexcvacket
    $. We get as in the NS case :
            \begin{equation}
              b_{n} \Gexcvacket
              = b^*_{n} \Gexcvacket
              = 0,
              \quad n \ge 1
              .
      \label{eq:generic_vacuum}
    \end{equation}

    \section{Asymptotic Fields and Relation Between Asymptotic
      Fields
      Vacua and the Vacuum}
    \label{sect:asymp_fields}
    
    In this section we define the asymptotic in-field and out-field
    and discuss how their vacua are related to that of the theory with
    defects.
    The relation is ``radial time dependent'' thus
    explicitly  showing that an interaction is hidden in the defects.
    In particular the vacuum for the theory with defects can be
    identified with $\SL$ in-field vacuum while it is connected by a
    Bogoliubov transformation to the $\SL$ out-field vacuum.
    
    In the following we use the expansion of
    \begin{equation*}
      P(z; \{x_t, \E t\} ) = \prod_{t=1}^N \qty( 1- \frac{z}{x_t} )^{E_t}
      ,
    \end{equation*}
    around the origin and infinity with coefficients   
    \begin{eqnarray*}
        \cmode{k}{0}{\E{t}}{x_t}
        &=&
          \sum_{ \{k_t\} \in \mathds{N}^N} 
          \prodt \qty[
          \binom{\E{t}}{k_t} \qty( - \frac{1}{x_t} )^{k_t}
          ]
          \updelta_{\sumt k_t, k}
      \\
        \cmode{k}{\infty}{\E{t}}{x_t}
        &=&
          \sum_{ \{k_t\} \in \mathds{N}^N} 
          \prodt \qty[
          \binom{\E{t}}{k_t} \qty( - x_t )^{k_t - \E{t}}
          ]
          \updelta_{\sumt k_t, k}
      , 
    \end{eqnarray*}
      so that we can write
    \begin{equation*}
      \begin{split}
        P(z; \{x_t, \E t\} )
        &      \underset{\abs{z} < x_N}{=}
          \sum_{k=0}^\infty \cmode{k}{0}{\E{t}}{x_t} z^k
          \nonumber\\
        &      \underset{\abs{z} > x_1}{=}
          \sum_{k=0}^\infty
          \cmode{k}{\infty}{\E{t}}{x_t} z^{-k+\sumt \E t}
          .
      \end{split}
    \end{equation*}
    We do not discuss intermediate fields, i.e. expansions for
    $x_t < \abs{z} < x_{t-1}$, as it is not possible to clearly disentangle
    the effects  of defects before and after this range
    since, as we will argue, the vacuum in presence  of
    defects is related to the radial ordering of the operators
    associated with the defects as in \eqref{eq:vacuum_R_prod_spin_fields}.
    
    \subsection{Asymptotic in-field
and relation between its vacuum and generic case vacuum}

Consider the definitions of the basis of solutions
\eqref{eq:generic-case-basis} and
\eqref{eq:generic-case-basis-conjugate} and expand around $z = 0$.
Let us concentrate on the first case since analogous relations can be
written for $b_n^{*\, ( 0 )}$ with the substitutions of $\E t $ with
$\bE t$.  We get for $0\le \abs{z} < x_N$
    \begin{eqnarray}
      \Uppsi_n( z ) &     \underset{\abs{z} < x_N}{=}  & \sum\limits_{k = 0}^{+\infty}
      \cmode{k}{0}{\E{t}}{x_t} \Uppsi^{( 0 )}_{n-k}( z ),
      \label{eq:expansion-around-0}
    \end{eqnarray}
    and
$
      \Uppsi^{( 0 )}_n( z ) = \normfac z^{-n}
$ as in \eqref{eq:usual-twisted-modes} with $\Es=0$ which are the
modes of a untwisted fermion, i.e. a plain NS fermion.
    The previous expansion connects the asymptotic
    behavior of the modes of the fermion with defects with the modes
    of a NS fermion which can be seen close to the origin.

    We can now relate the operators of the system with defects with
    those of the asymptotic in-field.
 To this purpose we can then substitute the expansion
    \eqref{eq:expansion-around-0} into the usual expression of the modes
    \eqref{eq:complex-plane-mode-expansion}:
    \begin{equation*}
      \Uppsi( z )
      =
      \sum\limits_{n \in \mathds{Z}} b_n \Uppsi_n( z )
      \underset{\abs{z} < x_N}{=}
      \Uppsi^{(in)}( z )
      =
      \sum\limits_{n \in
      \mathds{Z}} b_n^{( 0 )} \Uppsi_n^{( 0 )}( z )
    \end{equation*}
    thus leading to
    \begin{equation*}
      b_n^{( 0 )} = \sum\limits_{k = 0}^{+\infty} b_{n + k}
      \cmode{k}{0}{\E{t}}{x_t}.
    \end{equation*}
  
    By writing
    $
    \Uppsi^{( 0 )}_{n}( z )= \Uppsi_{n}( z )\,
        P(z; \{x_t, -\E t\} )
    $
    these expressions can also be inverted:
    \begin{equation*}
      \begin{split}
        b_n &
        =
        \sum\limits_{k = 0}^{+\infty}
        \cmode{k}{0}{-\E{t}}{x_t} b^{(0)}_{n + k},
      \end{split}
    \end{equation*}

  The important point is that annihilation operators of the asymptotic
  theory, i.e. operators with positive index,
  are expressed only using annihilation operators of the theory
  with defects, this means that we can set
  \begin{equation*}
    \Gexcvacket = \ket{0_{(in)}}_\SL.
    \end{equation*}

    \subsection{Relation between generic case vacuum and asymptotic
      out-field vacuum}
    As done in the previous section we can
    also explicitly compute the expansion
    for $\abs{z} > x_1$ (define for simplicity $\M=\sumt \E t$):
    \begin{eqnarray*}
      \Uppsi_n( z ) &     \underset{\abs{z} > x_1}{=} & \sum\limits_{k = 0}^{+\infty}
      \cmode{k}{\infty}{\E{t}}{x_t} \Uppsi^{( 0 )}_{n+k-\M}( z ),
    \end{eqnarray*}
    which connects the asymptotic behavior of the modes of the
    fermion with defects to the modes of a NS fermion which can be seen close
    to the infinity.

This relation can be used to link out-operators with the
operators of the theory with defects as
    \begin{equation*}
      \Uppsi(z)
      =
      \sum\limits_{n \in \mathds{Z}} b_n \Uppsi_n( z )
      \underset{|z|>x_1}{=}
      \Uppsi^{(out)}( z )
      =
      \sum\limits_{n \in
      \mathds{Z}} b_n^{( \infty )} \Uppsi_n^{( 0 )}( z )
    \end{equation*}
    thus getting
    \begin{equation}
      b_n^{( \infty )} = \sum\limits_{k = 0}^{+\infty} b_{n + \M - k}
      \cmode{k}{\infty}{\E{t}}{x_t}.
      \label{eq:b_inf-b}
    \end{equation}
    These expressions can also be inverted as
    \begin{equation*}
      \begin{split}
        b_n & = \sum\limits_{k = 0}^{+\infty} \cmode{k}{\infty}{-\E{t}}{x_t}
        b^{( \infty )}_{n + \M - k}.
      \end{split}
    \end{equation*}
    As we will show later, we must take $\M=0$.
    Then  the important point is that annihilation operators of the asymptotic
  theory, i.e. operators with positive index,
  are expressed using both annihilation and creator operators of the theory
  with defects while creators, i.e.  operators with non negative
  index,
  are expressed connected with creators only.
  It follows from the vacuum definition that
  \begin{equation*}
    \begin{split}
      &\qty( \cmode{0}{\infty}{-\E{t}}{x_t} b^{( \infty )}_{1 }
      +  \mbox{creators} ) \Gexcvacket =0
        \\
        &\qty( \cmode{0}{\infty}{-\E{t}}{x_t} b^{( \infty )}_{2 }
          + \cmode{1}{\infty}{-\E{t}}{x_t} b^{( \infty )}_{1 }
          +  \mbox{creators} ) \Gexcvacket =0
          \\
          &\vdots
        \end{split}
        ,
    \end{equation*}
  This means that the vacuum for the asymptotic out-field is non
  trivially connected to the vacuum of the theory with defects.
  More explicitly we get the relation
    \begin{equation*}
      \Gexcvacket = {\cal N}_{(out)}(\{x_t, \E t,\bE t\})
      e^{\sum_{m,n\le 0}
        {\cal M}_{m n}^{(out)}(\{x_t, \E t, \bE t\})
        b^{( \infty ) *}_{m } b^{( \infty )}_{n } }
      \ket{0_{(out)}}_\SL
      ,
    \end{equation*}
    so that the two $\SL$ vacua are connected by a Bogoliubov transformation.
    More precisely  we get (see appendix \ref{app:details_reflection}
    for details)
    \begin{align}
      \qty{
            \Uppsi^{(out, +)}( z )
      +
        \oint_{|z|,|w|> x_1} \frac{d w}{ 2\uppi i}
        \frac{ P(z; \{x_t, \E t\} ) P(w; \{x_t, -\E t\} ) -1
        }{
        z-w}
        \Uppsi^{(out, -)}( w )
        }
      \Gexcvacket = 0
      ,
      \label{eq:reflection condition_out_field_generic_vacuum}
      \end{align}
      and the corresponding equation for $\Uppsi^{(out) *}( z )$
      with the substitution $\Es  \rightarrow \bEs$.
      Notice that the kernel of the integral is nothing else (up to a
      multiplicative constant) but the
      regularised propagator, i.e. the propagator in the presence of
      defects \eqref{eq:gen_Radial_order} to which the NS propagator has been
      subtracted. 
      The previous equation can be solved explicitly by
      \begin{equation*}
        \begin{split}
          \Gexcvacket
          =&
             {\cal N}(\{x_t, \E t,\bE t\})
          \\
          &
             e^{
            \oint_{|z|,|w|> x_1}
            \frac{d z}{ 2\uppi i}  \frac{d w}{ 2\uppi i}
          \Uppsi^{(out, -) *}( z )
            \frac{ P(z; \{x_t, \E t\} ) P(w; \{x_t, -\E t\} ) -1
          }{
          z-w}
          \Uppsi^{(out, -)}( w )
        }
        \ket{0_{(out)}}_\SL
        .
        \end{split}
      \end{equation*}
      In the previous equation there is no need to
      specify whether $\abs{z}$ is greater or less than $\abs{w}$ since
      $\Uppsi^{(out, -)*}( z )$ and $\Uppsi^{(out, -)}( w )$
      anti-commute.
      Finally deriving the same expression using
      $\Uppsi^{(out, -)*}( z )$ and comparing with the previous one
      we deduce that $\bE t = -\E  t$.
      
    
  \section{Contractions and Stress-Energy Tensor}
  \label{sec:contraction_and_T}

  Given the definitions of the in-vacuum of the theory and the algebra of
  operators, we can finally define the normal ordering operation and
  proceed to compute the contractions and OPEs of the operators: the procedure
  ultimately leads to the definition of the stress-energy tensor. This is enough
  to show that the theory is a time dependent CFT since the stress-energy tensor
  satisfies the canonical OPE.

    \subsection{NS Complex Fermion}

    First of all we deal with the simple case of NS fermions and using the
    algebra \eqref{eq:ns-algebra} we compute the OPE of fermion
    fields 
    \begin{equation*}
      \Uppsi^i( z ) \Uppsi^*_j( w )  = \no{\Uppsi^i( z ) \Uppsi^*_j( z
      )} + \frac{1}{\uppi T} \frac{\updelta^i_j}{ z - w },\quad \abs{w} <
      \abs{z},
    \end{equation*}
    where the operation $\no{\cdot}$ is the normal ordering with
    respect to the $\SL$ vacuum defined in \eqref{eq:NS_SL2_vacuum} .

    Secondly we get to the expression of the stress-energy tensor:
    \begin{equation*}
      \begin{split}
        \mathcal{T}( z ) & = \lim\limits_{w \to z} \qty[
        -\frac{ \uppi T }{2}
        \qty( \Uppsi^*_i( z ) \partial_w \Uppsi^i( w ) - \partial_z \Uppsi^*_i(
        z ) \Uppsi^i( w ) )
        + \frac{\Nf}{ ( z - w )^2 } ]
        \\
        & = -\frac{\uppi T}{2} \no{ \Uppsi^*_i( z ) \lrpartial{z} \Uppsi^i( z ) }
      \end{split}
    \end{equation*}
    so we are now able to derive the necessary minimal subtraction  
    \begin{equation*}
      \mathfrak{h}( z - w ) =  \frac{\Nf}{ ( z - w )^2 }
      ,
    \end{equation*}
    to get the stress-energy tensor.

    \subsection{Twisted Fermion}

    We can now go back to $\Nf = 1$ theories. First of all we consider the
    simplest case of the usual twisted fermion with the mode expansion
    \eqref{eq:usual-twisted-mode-expansion} and
    \eqref{eq:usual-twisted-mode-expansion-conjugate}.
    We do not implement beforehand the constraint
    \eqref{eq:twisted-fermion-consistency} but we recover it in a
    different way.
    Both excited and twisted vacua can be treated on the same footing
    since the their difference amount to choose $n_{\Es}$ and $n_{\bEs}$.

      \subsubsection{OPE and Stress-Energy Tensor}

      Using the anti-commutation relations
      \eqref{eq:twisted-fermion-algebra} we can compute the OPE

      \begin{equation*}
        \Uppsi( z ) \Uppsi^*( w ) = \noE{ \Uppsi( z ) \Uppsi^*( w ) } +
        \frac{1}{\uppi T} \qty( \frac{z}{w} )^{\Es} \frac{1}{z - w},
        \quad
        \abs{w} < \abs{z},
      \end{equation*}
      and
      \begin{equation*}
        \Uppsi^*( w ) \Uppsi(z ) = \noE{ \Uppsi^*( w ) \Uppsi( z ) } +
        \frac{1}{\uppi T} \qty( \frac{w}{z} )^{\bEs} \frac{1}{w - z},
        \quad
        \abs{w} > \abs{z}.
      \end{equation*}
      If we require that the previous results can be assembled in a
      well defined continuous radial ordering
      $ R\qty[ \Uppsi( z ) \Uppsi^*( w ) ]
      $ we need to set $\Es=-\bEs$ so we can write
      \begin{equation*}
        R\qty[ \Uppsi( z ) \Uppsi^*( w ) ]
        =
        \noE{ \Uppsi( z ) \Uppsi^*( w ) } +
        \frac{1}{\uppi T} \qty( \frac{z}{w} )^{\Es} \frac{1}{z - w}
        .
        \end{equation*}

      The same result can be reached by computing the
      stress-energy tensor starting from the previous
      expressions. We have two ways to construct it depending on the
      ordering of the classical expression.
      Either as
      \begin{equation}
        \begin{split}
          \mathcal{T}( z ) & = \lim\limits_{\substack{w \to z \\ \abs{w} <
              \abs{z}}}
          \qty[
          -\frac{\uppi T}{2} \qty(
          \Uppsi^*( z ) \partial_w \Uppsi( w )
          - \partial_z \Uppsi^*( z ) \Uppsi( w ) )
          + \frac{1}{( z- w )^2} ]
          \\
          & =
          -\frac{\uppi T}{2} \no{ \Uppsi^*( z ) \lrpartial{z} \Uppsi( z ) }
          +\frac{\Es^2}{2 z^2},
        \end{split}
        \label{eq:T_excited_vacuum}
      \end{equation}
      or
      \begin{equation*}      
        \begin{split}
          \mathcal{T}( z ) & = \lim\limits_{\substack{w \to z \\ \abs{w} <
              \abs{z}}}
          \qty[ -\frac{\uppi T}{2}
          \qty(
          - \partial_z \Uppsi( z ) \Uppsi^*( w )
          +\Uppsi( z ) \partial_w \Uppsi^*( w )
          )
          + \frac{1}{( z -  w )^2} ]
          \\
          & =
          -\frac{\uppi T}{2} \no{ \Uppsi^*( z ) \lrpartial{z} \Uppsi( z ) }
          +\frac{\bEs^2}{2 z^2},
        \end{split}
      \end{equation*}
      which however must coincide for consistency. Since
      \begin{equation*}
        \no{ \Uppsi( z ) \lrpartial{z} \Uppsi^*( z ) } = \no{ \Uppsi^*( z )
        \lrpartial{z} \Uppsi( z ) } \,,
      \end{equation*}
      then we must then require
$
        \Es^2 = \bEs^2 $.
 
      We can  get a stronger constraint by computing the OPE
      $\mathcal{T}( z ) \mathcal{T}( w )$.
      In fact the cancellation of the cubic divergence requires $\Es+\bEs=0$.

      It the follows that the vacuum $\excvacket$ is actually
      $\eexcvacket$, notation we will use from now on.


      \subsubsection{Virasoro Operators and Conformal Dimensions}

      From the usual definition of the
      stress-energy tensor in terms of the Virasoro generators
$
        \mathcal{T}( z ) = \sum\limits_{k \in \mathds{Z}} L_k z^{-k-2},
$
      we can extract the operators $L_k$ from any of the previous definitions:
      \begin{equation*}
        \begin{split}
          L_{(\Es) k}
          &=
        -\frac{\uppi T}{2} \normfac^2 \sum\limits_{n \in \mathds{Z}} \noE{
          \obbE_n \obE_{k + 1 - n} } ( 2n - k + 2 \Es - 1 ) 
        + \frac{\Es^2}{2} \updelta_{k,0}
        \\
        &=
        \frac{\uppi T}{2} \normfac^2 \sum\limits_{n=1}^{\infty}
        \Big[
            ( 2n - k + 2 \Es - 1 ) 
            \noE{  \obE_{k + 1 - n}  \obbE_n}
            \\
            &+
          ( 2n - k - 2 \Es - 1 ) 
          \noE{  \obbE_{k + 1 - n}     \obE_n} 
        \Big]
          + \frac{\Es^2}{2} \updelta_{k,0}
      \end{split}
.
    \end{equation*}

    Looking back at the analysis of the excited and twisted vacua, we already
      hinted to the fact that they are not in general $\SL$ invariant. 
      In particular we can see that that the excited vacua  $\eexcvacket$
      is a primary field
      \begin{equation*}
        L_{(\Es) k > 0} \eexcvacket =0,
        \qquad
        L_{(\Es) 0} \eexcvacket = \frac{\Es^2}{2} \eexcvacket
        ,
        \end{equation*}
      with non trivial conformal dimensions
      $\Updelta\qty( \eexcvacket ) =  \frac{\Es^2}{2}$.
      This operator is an excited spin field $\spin{\E{t}}{x}$
      inserted at $x=0$ whose bosonized expression is given by
      \begin{equation*}
        \spin{\Es}{x} = e^{i \Es \upphi( x )},
      \end{equation*}
      where $\upphi$ is such that
      \begin{equation*}
        \left\langle \upphi( z ) \upphi( w ) \right\rangle = -\frac{1}{( z -
          w)^2}
        .
      \end{equation*}

      In fact the minimal conformal dimension is achieved
      for $n_{\Es}=n_{\bEs}=0$, i.e.
      $
      \Updelta\qty( \twsvacket ) =  \frac{\epss^2}{8}
      $
      and we know this is the basic spin field.
      We can further check this idea 
      by showing that the conformal dimensions are consistent.
      Using \eqref{eq:usual-twisted-fermion-conformal-twisted} we get
      \begin{equation*}
        \begin{split}
          L_{(\Es) 0}\twsvacket
          & =
          L_0
          \qty( \obbE_0 \obbE_{-1} \dots \obbE_{2-n_{\Es}}   \eexcvacket)
          \\
          & =
          \left[ \sum\limits^{n_{\Es}}_{n = 1} ( n - \frac{\Es +  1}{2} )
            +\frac{\Es^2}{2}
            \right] \twsvacket
           = +\frac{1}{8} \epss^2\twsvacket .
        \end{split}
      \end{equation*}

    \subsection{Generic Case With Defects}

    We will now apply the same procedure to the generic case of one complex
    fermion in the presence of an arbitrary number of spin fields with respect
    to the vacuum we introduced in \eqref{eq:generic_vacuum}. We will consider the mode expansion
    \eqref{eq:generic-case-basis} and \eqref{eq:generic-case-basis-conjugate} as
    well as the anti-commutation relations
    \eqref{eq:generic-case-anti-commutation}.

      As in the usual twisted case, we will first consider the contraction of
      the field $\Uppsi$ and $\Uppsi^*$ and then move to the stress-energy
      tensor. Using the anti-commutation relations
      and $\sum\limits_{k \in \mathds{Z}} p_k z^k = \prodt \qty( 1 - \frac{z}{x_t} )^{-\LL{t}}$
      where $p_k$ is defined in \eqref{eq:generic-conserved-product-factor}.
      We have:
      \begin{equation*}
        \Uppsi( z ) \Uppsi^*( w ) = \no{ \Uppsi( z ) \Uppsi^*( w ) } +
        \frac{1}{\uppi T} \frac{1}{z - w} \prodt \qty( 1 - \frac{z}{x_t}
        )^{\E{t}} \qty( 1 - \frac{w}{x_t} )^{-\E{t}},
      \end{equation*}
      as well as
      \begin{equation*}
        \Uppsi^*( z ) \Uppsi( w ) = \no{ \Uppsi^*( z ) \Uppsi( w ) } +
        \frac{1}{\uppi T} \frac{1}{z - w} \prodt \qty( 1 - \frac{z}{x_t}
        )^{\bE{t}} \qty( 1 - \frac{w}{x_t} )^{-\bE{t}},
      \end{equation*}
      both for $\abs{w} < \abs{z}$.
      If we require that the previous results can be assembled in a
      well defined continuous radial ordering
      $ R\qty[ \Uppsi( z ) \Uppsi^*( w ) ]
      $ we need to set $\E t=-\bE t$ so we can write
      \begin{equation}
        R\qty[ \Uppsi( z ) \Uppsi^*( w ) ]
        =
        \no{ \Uppsi( z ) \Uppsi^*( w ) } +
        \frac{1}{\uppi T} \frac{1}{z - w}
        \prodt \qty( 1 - \frac{z}{x_t})^{\E{t}}
        \qty( 1 - \frac{w}{x_t} )^{-\E{t}}
        .
        \label{eq:gen_Radial_order}
        \end{equation}
      We can then expand the results around $z$:
      \begin{eqnarray*}
        \begin{aligned}
          R\qty[\Uppsi( z ) \Uppsi^*( w )]
          & =
          \no{ \qty(\Uppsi\Uppsi^*)( z ) }
          +
          \no{ \qty(\Uppsi\partial\Uppsi^*)( z ) }\, (w-z)
          \\
          & + \frac{1}{\uppi T} \left[
            \frac{-1}{w- z}
            + \sumt \frac{\E{t}}{z - x_t} \right.
          \\
          & \left.
            - \frac{1}{2} \qty(
            \sumt \sum\limits_{u \neq t}
              \frac{\E{t} \E{u}}{( z - x_t) ( z - x_u )}
            + \sumt \frac{\E{t} \qty( \E{t} - 1 )}{( z - x_t )^2} )
            ( w - z )
          \right]
          \\
          &  + \order{( w - z )^2}
          ,
        \end{aligned}
      \end{eqnarray*}
      and around $w$
      \begin{eqnarray*}
        \begin{aligned}
          R\qty[\Uppsi( z ) \Uppsi^*( w )]
          & =
          \no{ \qty(\Uppsi\Uppsi^*)( w ) }
          +
          \no{ \qty(\partial\Uppsi\Uppsi^*)( w ) }\, (z-w)
          \\
          & + \frac{1}{\uppi T} \left[
            \frac{1}{z- w}
            + \sumt \frac{\E{t}}{w - x_t} \right.
          \\
          & \left.
            + \frac{1}{2} \qty(
            \sumt \sum\limits_{u \neq t}
              \frac{\E{t} \E{u}}{( w - x_t) ( w - x_u )}
            + \sumt \frac{\E{t} \qty( \E{t} - 1 )}{( w - x_t )^2} )
            ( z - w )
          \right]
          \\
          &  + \order{( z - w )^2}
          ,
        \end{aligned}
      \end{eqnarray*}      
      so that the stress-energy tensor      becomes:
      \begin{align*}
          \mathcal{T}( z ) & = -\frac{\uppi T}{2} \no{ \Uppsi( z ) \lrpartial{z}
          \Uppsi^*( z ) } 
        + \frac{1}{2} \qty( \sumt  \frac{\E{t}}{z - x_t})^2
                             \nonumber\\
        &=
          \frac{\uppi T }{2} \normfac^2
          \sum_{n,m} : b_n b^*_m:
          z^{-n-m}
          \qty[ \frac{m-n}{z}+2\sumt \frac{\E t}{z-x_t} ]
        + \frac{1}{2} \qty( \sumt  \frac{\E{t}}{z - x_t})^2
          .
      \end{align*}
      The last expression shows that the energy momentum tensor $\mathcal{T}( z )$
      is radial time dependent but it satisfies the usual OPE.      

      Notice first of all that the vacuum $\Gexcvacket$ is actually
      $\GGexcvacket$, i.e. it depends only on $x_t$ and $\E t$.
      Then we can try to interpret the previous result in the light of the
      usual CFT approach.
      In particular we can refine the idea we discussed after
      \eqref{eq:asymp_beha_Psi_on_exc_vac} that
      the singularity in the modes \eqref{eq:generic-case-basis} and
      \eqref{eq:generic-case-basis-conjugate}
      at the point $x_t$
      is associated with a primary conformal
      operator which creates $\eexcvacket$ with $\Es=\E t$.
      In fact  by comparison with the stress energy tensor of a
      excited vacuum
      \eqref{eq:T_excited_vacuum},
      we can read from the second order singularity
      that at the points $x_t$ there is an operator
      which creates the excited vacuum $\GGexcvacket$ from the
      $\SL$ vacuum ${\ket 0}_\SL$.
      Given the discussion in the previous section this is an excited
      spin field $\spin{\E{t}}{x_t} = e^{i \E{t} \upphi( x_t )}$.
      The first order singularities in $x_u-x_t$ are then the result
      of the interaction between two of the previous excited spin
      fields.
      We can try to be more precise.
      Using the usual CFT operatorial approach
      we can suppose that the following
      identification holds
      \begin{align}
        \GGexcvacket
        &=
        {\cal N}(\{x_t, \E t\})~
        \spin{\E{1}}{x_1} \dots \spin{\E{N}}{x_N}{\ket 0}_\SL
          \nonumber\\
        &=
        {\cal N}(\{x_t, \E t \})~
        R\qty[ \prodt \spin{\E{t}}{x_t} ] {\ket 0}_\SL
        ,
          \label{eq:vacuum_R_prod_spin_fields}
      \end{align}
      then we get
      \begin{align*}
          \mathcal{T}( z )         \GGexcvacket
        &=
        {\cal N}(\{x_t, \E t \})~
          R\qty[
          \mathcal{T}( z )
          \prodt \spin{\E{t}}{x_t} ] {\ket 0}_\SL
          .
      \end{align*}
      The fact that $ \mathcal{T}( z )$ enters the radial ordering may
seem strange but the left hand side is well defined for all $z$ and
the only well defined expression for the right hand side is the one
with the radial ordering.
In fact an operatorial expression like
$\mathcal{T}( z ) R\qty[\partial\upphi(x_1)\partial\upphi(x_2)]{\ket
  0}_\SL$
is only defined for $|z|> x_{1,2}$.
      It then follows that
    \begin{equation*}
          \mathcal{T}( z )         \GGexcvacket
       =
          \sumt \qty(
          \frac{\E{t}^2 /2 }{ (z-x_t)^2}
          +
          \frac{\ \partial_{x_t} - \partial_{x_t} \log {\cal N} }{ z-x_t}
          ) \GGexcvacket
          + \mbox{regular terms in $z$}
          ,
      \end{equation*}
which allows to write
\begin{align*}
        {\cal N}(\{x_t, \E t \})~
          &R\qty[
          \partial_{x_t} \spin{\E{t}}{x_t} ]
          \prod_{u\ne t} \spin{\E{u}}{x_u} ] {\ket 0}_\SL
            \nonumber\\
            &=
          \E t\,
          \qty[
          \uppi T \normfac^2\,  
          \sum_{n,m=0}^\infty \frac{ b_n b^*_m}{x_t^{n+m}}
          +
          \sum_{u\ne t} \frac{\E{u}}{x_t-x_u}
          ]
              \GGexcvacket
              .
\end{align*}        
This result shows the way non primary operators are reresentend in
this formalism and is consistent with the computation of the excited
spin fields correlator performed in section \ref{sec:spin_correlators}.

  \section{Hermitian Conjugation}
  \label{sec:hermitian_and_outvacuum}

  In this section we focus on the operation of ``Hermitian
  conjugation''.
  We write ``Hermitian conjugation'' between quotes because the
  Hermitian conjugation requires the existence of an inner
  product which is not yet available
  since we have not defined the out-vacuum.
  What we are going to discuss is actually more like the involutive
  $\star$ operator of $C^\star$ algebras with the catch that the
  $\star$ operator sends an element of an algebra to another
  element of the same algebra.
  This is not what happens in the generic case since the $\star$ is
  essentially associated with the inversion $z \rightarrow
  \frac{1}{\bar z}$, i.e. in evolving from $\uptau=+\infty$ to
  $\uptau=-\infty$ so that the order of boundary singularities is
  reversed.
  In the next section we use this correspondence between the $\star$
  operator we defined and the Hermitian conjugation to define the out-vacuum.
  The previous warning does not apply to the usual twisted fermions
  with  which we start.
  
    \subsection{Usual Twisted Fermions}
    In general for a chiral primary conformal operator of dimension
    $\Delta$ in $z$ coordinates the Euclidean Hermitian conjugation is
    \begin{equation*}
      \qty[ O(z) ]^\dagger = \eval{ \qty( w^{2\Delta} O(w) ) }_{w=1/ \bar z}.
    \end{equation*}
    As discussed above  we cannot use the words
    ``Euclidean Hermitian conjugation'' and
    define it  since we do not have an inner product 
    but we can define the operation 
    $\star$ which mimics its  behavior.
    Therefore we define
    \begin{equation}
        \qty[ \Uppsi( z; \Es ) ]^{\star}
        = \eval{
          \qty[ w\, \widetilde{\Uppsi}^*( w; -\widetilde \Es ) ]
        }_{w=1/ \bar z},
        \quad
        \qty[ \Uppsi^*( z; \Es ) ]^{\star}
        = \eval{
            \qty( w\, \widetilde{\Uppsi}( w; \widetilde \Es ) )
        }_{w=1/ \bar z},
      \label{eq:star_on_Psi}
    \end{equation}
      where we have not assumed that the action of $\star$ is a map between
      the same space and we have written for example $ \Uppsi( z; \Es )$
      to make explicit the dependence on the parameter $\Es$ which
      enters in the modes.
      The previous action agrees with \eqref{eq:complex-plane-conjugate}.
      In terms of the basis \eqref{eq:usual-twisted-modes}, we can
      write\footnote{
        The other possibility
        $\qty[ \Uppsi_n^{( \Es )}(z) ]^{\star}
         =
         \eval{
           \qty[ w\, \Uppsi_{-n}^{*\, ( -\Es-1 )}\qty( w ) ]
         }_{w=1/ \bar z}$ is inconsistent with the anti-commutation relations.
      }
    \begin{equation*}
      \qty[ \Uppsi_n^{( \Es )}(z) ]^{\star}
      =
      \eval{
        \qty[ w\, \Uppsi_{1-n}^{( -\Es )\, *}\qty( w ) ]
      }_{w=1/ \bar z},
      \qquad
      \qty[ \Uppsi_n^{( -\Es )}(z) ]^{\star}
      =
      \eval{
        \qty[ w\, \Uppsi_{1-n}^{( \Es )\, *}\qty( w ) ]
      }_{w=1/ \bar z},
    \end{equation*}
      which shows that in this case the image of the $\star$ operator is the
      same of the support.
      Using the mode expansion of \eqref{eq:star_on_Psi} it follows that
      \begin{equation}
      \qty[ \obE_n ]^{\star} = \obbE_{1-n},
      \qquad
      \qty[ \obbE_n ]^{\star} = \obE_{1-n}.
      \label{eq:star_usual_twisted}
    \end{equation}
    The $\star$ action is compatible with the anti-commutation relations
    as we can show by explicitly computing them:
    \begin{equation*}
      \qty( \qty[ \obE_n, \obbE_m ]_+ )^{\star} = \qty[ \obbE_{1-n},
      \obE_{1-m} ]_+ = \frac{1}{\uppi T \normfac^2} \updelta_{n+m,1}.
    \end{equation*}
    Furthermore $\star$ is involutive since:
    \begin{equation*}
      \qty[ \Uppsi_n^{( \Es )}( z ) ]^{\star \star} 
      = \Uppsi_n^{( \Es )}( z ) \quad
      \Rightarrow \quad \qty[ \obE_n ]^{\star \star} = \obE_n.
    \end{equation*}
%
    \subsection{Generic Case With Defects}

    The situation in the generic case is more complex.
    Consider the modes given in \eqref{eq:generic-case-basis}
    then it is natural to
    define the action of the $\star$ operator on them as:
    \begin{equation*}
      \begin{split}
        \qty[ \Uppsi_{n} ( z; \{x_t, \E t\} ) ]^{\star}
        & =
        \normfac\,
        \overline{z}^{-n} \prodt \qty( 1 - \frac{\overline{z}}{x_t} )^{\E{t}}
        \\
        & =
        \eval{ \qty(
        w \, \prodt \qty( - \frac{1}{x_t} )^{\E{t}} \,
        \normfac\,
        w^{-( M + 1 - n )}
        \prodt \qty( 1 - \frac{w}{ 1/ x_t}  )^{\bE{t}}
        ) }_{w=1/ \bar z}
        \\
        & =
        \eval{ \qty(
        w \, \prodt \qty( - \frac{1}{x_t} )^{\E{t}} \,
        \widetilde{\Uppsi}_{M + 1 - n}^*
        \qty(w; \{\widetilde{x}_t, \btE t\})
        ) }_{w=1/ \bar z}
      \end{split}
    \end{equation*}
    where we used $\M = \sumt \E{t}$.
In this case
    the image of the $\star$ operator is a
    different space where the defects are located in $\widetilde{x}_t$
    and the singularities are $\tE t$ and $\btE t$ with
    \begin{equation*}
      \widetilde{x}_t = \frac{1}{x_t},
      \qquad \tE{t} = -\E{t},
      \qquad \btE{t} = \E{t},      
    \end{equation*}
    where we used $\E{t} + \bE{t} = 0$.

    We can therefore compute the action of the $\star$ operator on the
    creation and annihilation operators as done previously and get:
    \begin{equation*}
      b_n^{\star} = \prodt \qty( - \frac{1}{x_t} )^{-\E{t}} \,
      \widetilde{b}^*_{{M} + 1- n}
      ,
      \qquad
      (b_n^*)^{\star} = \prodt \qty( - \frac{1}{x_t} )^{\E{t}} \,
      \widetilde{b}_{-M + 1- n}
      .
    \end{equation*}
    As in the previous situation, the anti-commutation relations are
    preserved by the $\star$ operator. Explicitly we have:
    \begin{equation*}
      \qty( \qty[ b_n, b_m^* ]_+ )^{\star}
      = \qty[ 
      \widetilde{b}_{-M + 1 - m}, 
      \widetilde{b}_{{M} + 1 - n}^* ]_+
      = \frac{1}{\uppi T \normfac^2}
      \updelta_{n + m, 1}
      .
    \end{equation*}
    Finally the $\star$ operator is involutive.

  \section{Definition of the Out-Vacuum}
  \label{sec:out-vacuum}
  With the definition of the $\star$ operator we can now
  proceed to define the out-vacuum such that it
  acts as the Hermitian conjugation in the usual cases.
  It is conceptually separated from the definitions
  of the algebra of operators and their representation on the
  in-vacuum.
  This is the last step
  before we can compute any correlation function.
  We first consider the usual twisted theory from which we
  learn how to define the out-vacuum and then move to the
  generic case in the presence of multiple defects.

    \subsection{Usual Twisted Fermions}

    Consider the definition of the in-vacuum
    \eqref{eq:usual-excited-vacuum} for the fields image of the
    $\star$ operator, i.e. 
    $\widetilde \Uppsi( w;\widetilde \Es )$ and
    $\widetilde \Uppsi^*( w; \btEs )$.
    It is defined as
    \begin{equation*}
      {\widetilde b}^{( \tEs )}_n \ket{\widetilde T_{\tEs ,\btEs}}
      = 
      {\widetilde b}^{( \btEs ) *}_n \ket{\widetilde T_{\tEs ,\btEs}}
      = 0,
      \quad n \ge 1.
    \end{equation*}
    Then the usual Hermitian conjugation gives
    \begin{equation*}
      \bra{\widetilde T_{\tEs ,\btEs}} \qty( {\widetilde b}^{( \tEs )}_n)^\dagger 
      = 
      \bra{\widetilde T_{\tEs ,\btEs}} \qty( {\widetilde b}^{( \btEs ) *}_n)^\dagger 
      = 0,
      \quad n \ge 1.
    \end{equation*}

    Given the action of the $\star$ operator
    \eqref{eq:star_usual_twisted},
    if we want to identify it with the Hermitian conjugate we
    are led to write
    \begin{equation*}
      \eexcvacbra \obE_n = \eexcvacbra \obbE_n = 0, \quad n \le 0.
    \end{equation*}

    \subsection{Generic Case With Defects}

    We can now analyze the case of an arbitrary number of defects using previous
    relations.
    Following the steps of the previous section we can define the
    in-vacuum for the tilded theory as
        \begin{equation*}
      {\widetilde b}_n \GGexcvacket
      = 
      {\widetilde b}_n \GGexcvacket
      = 0,
      \quad n \ge 1,
    \end{equation*}
    and then interpret it as the out-vacuum for the initial theory.     
    The definition of the out-vacuum is therefore:
      \begin{eqnarray*}
         \GGexcvacbra b_n & = & 0, \quad n \le \M,
        \\
        \GGexcvacbra b^*_n & = & 0, \quad n \le - \M.
      \end{eqnarray*}
      Since the action of the $\star$ operator is compatible with the
      anti-commutation relations this definition is consistent as
      definition for the out-states.
      The consistency between the in-vacuum and out-vacuum is however not
      granted and must be checked.
      If we assume that
      $
      \braket{\GGexcvac}{\GGexcvac}
      \ne 0 $
      then using the anti-commutation relations we get
      \begin{equation*}
        \begin{split}
          \frac{1}{\uppi T \normfac^2} \braket{\GGexcvac}{\GGexcvac}
          &=
            \GGexcvacbra
            \qty[ b_{\M}, b_{-\M+1}^* ]_+
            \GGexcvacket
          \\
          &=
            \GGexcvacbra
            b_{-\M+1}^*  b_\M
            \GGexcvacket
          \ne 0,
        \end{split}
      \end{equation*}
      which requires
      $b_\M
      \GGexcvacket
      \ne 0
      $
      There is a  similar condition for the $b^*_{-\M}$, therefore we
      must require $\M \le 0$ and $-\M \le 0$, thus
      \begin{equation*}
       \M =\sumt \E{t} = 0.
      \end{equation*}
      The situation is therefore analogous to the case depicted in
      Figure~\ref{fig:inconsistent-theories} where $\M$ and $\bM$ have the same
      role of $\LLs$ for the twisted fermion.

      \subsection{Asymptotic vacua}
      The discussion is essentially the same as in section
      \ref{sect:asymp_fields} with the role of asymptotic in- and out-fields
      exchanged.
      In particular we get
      \begin{equation*}
        \GGexcvacbra = {}_\SL\bra{0_{(out)}}
        ,
     \end{equation*}
     and
     \begin{equation*}
      \GGexcvacbra = {\cal N}_{(in)}(\{x_t, \E t\})
      ~~{}_\SL\bra{0_{(in)}}
      e^{\sum_{m,n\ge 1}
        {\cal M}_{m n}(\{x_t, \E t\})
        b^{(0 ) *}_{m } b^{( 0 )}_{n } }
      .
    \end{equation*}

  \section{Spin Field Correlators}
  \label{sec:spin_correlators}
      
  The definitions of the in- and out-vacua and the stress-energy tensor are
  critical to compute any correlation function of operators in the presence of
  the point-like defects. In fact we need to know both the algebra of the
  operators and their representation, usually defined on the in-vacuum (the ket
  vector), as well as their Hermitian conjugation in order to build the action
  of the operators on the out-vacuum (the bra vector).

  Starting from \eqref{eq:vacuum_R_prod_spin_fields}
  we can finally compute the spin field correlators
  \begin{equation*}
    \braket{\GGexcvac}
        =
        {\cal N }(\{x_t, \E t \})
        \left\langle
          R\qty[
          \prodt \spin{\E{t}}{x_t}
          ]
        \right\rangle
        .
  \end{equation*}
  At first sight this expression might look incorrect since both
  $\GGexcvacket$  and $\GGexcvacbra$ seem to contain 
  $ R\qty[ \prodt \spin{\E{t}}{x_t} ] $ as if were squaring
  the previous radial ordering.
  That it is not the case and it can be seen in different ways.
  The simplest is to realize that such a square would be
  divergent while the product seems to be perfectly finite.
  A more sophisticated and rigorous way is to consider what the
  previous product is from the point of view of asymptotic out field.
  In this case
  $\GGexcvacket= {\cal N}_{(out)}~ R\qty[ \prodt \spin{\E{t}}{x_t} ] \ket{ 0_{(out)}}_\SL$
  and $\GGexcvacbra= {}_\SL\bra{ 0_{(out)}}$ so that
  $
  {\cal N}_{(out)}={\cal N}
  $.
  Moreover $T(z) \underset{\abs{z} >x_1}{=} T_{(out)}(z)$ when the two
  energy momentum tensors are normal ordered with respect to their
  different sets of operators which are related as in \eqref{eq:b_inf-b}.
  Hence all the expressions are surely valid for $\abs{z} >x_1$ and can
  be analytically extended to the whole plane. 
  The same result can be obtained from the point of view of asymptotic
  in-fields.

  Unfortunately it is not completely clear how to fix the
  normalization.
  Moreover the result depends on the normalization chosen for the
  single spin field and this normalization shows only up when we
  relate the $N$ points correlators to $N-1$ points ones and these
  recursively down to two points correlators.    
  Therefore we need to consider quantities where the
  normalization cancels.
  In particular we can consider
  \begin{equation*}
    \begin{split}
      & \pdv{x_{\myoverline{t}}}
      \ln \left\langle
        R\qty[
        \spin{\E{\myoverline{t}}}{x_{\myoverline{t}}}
        \prod\limits_{u = 1,
          u \neq \myoverline{t}}^N \spin{\E{u}}{x_u}
        ]
      \right\rangle
      \\
      &=
      \oint\limits_{\abs{z} = x_{\myoverline{t}}} \frac{\dd{z}}{2\uppi        i}
      \frac{
        \left\langle
          R\qty[
          \mathcal{T}( z )
          \prodt \spin{\E{t}}{x_t}
          ]
        \right\rangle
      }{
        \left\langle
          R\qty[
          \prodt \spin{\E{t}}{x_t}
          ]
        \right\rangle
      }
        \\
        & =
        \qty( \oint\limits_{\abs{z} > x_{\myoverline{t}}} \frac{\dd{z}}{2\uppi
      i} - \oint\limits_{\abs{z} < x_{\myoverline{t}}} \frac{\dd{z}}{2\uppi i}
    )
    \frac{\GGexcvacbra \mathcal{T}( z )
      \GGexcvacket}{\braket{ \GGexcvac }
      }
      \\
      & = 
      \frac{\GGexcvacbra \qty(
      L_{-1}^{x_{\myoverline{t}}^+} - L_{-1}^{x_{\myoverline{t}}^-} )
      \GGexcvacket
    }{
      \braket{\GGexcvac }
        }
    \end{split}
    ,
  \end{equation*}
  since
$   
      \qty[ L_{-1}, \mathcal{O}_h( z ) ] = \partial_z \mathcal{O}_h( z )
 $
    for a quasi-primary operator $\mathcal{O}_h$.
 From the definition of $\mathcal{T}( z )$ it follows that:
  \begin{equation*}
    L_{-1}^{x_t^+} - L_{-1}^{x_t^-}
    =
    \oint\limits_{\mathcal{C}_{x_t}}
    \frac{\dd{z}}{2 \uppi i} \mathcal{T}( z )
    =
    \uppi T\, \normfac^2\,
    \E t\,
    \sum_{n,m} : b_n b^*_m: x_t^{-m-n}
    +
    \sum\limits_{u = 1, u \ne t}^N
    \frac{\E{u} \E{t}}{x_t - x_u},
  \end{equation*}
  where $\mathcal{C}_{x_t}$ is a small path circling $x_t$. Therefore
  \begin{equation*}
    \pdv{x_{\myoverline{t}}} \ln \left\langle
      R\qty[
      \prod\limits_{u} \spin{\E{u}}{x_u}
      ]
    \right\rangle
  =
   \sum\limits_{u \neq t} \frac{\E{u} \E{t}}{x_t - x_u},
  \end{equation*}
  which can be solved by
  \begin{equation*}
    \left\langle R\qty[ \prodt \spin{\E{t}}{x_t} ] \right\rangle =
    \mathcal{N}_0\qty( \{\E t\} )
    \prod\limits_{t=1,t >u}^{N} \qty( x_u - x_t )^{\E{u}
    \E{t}}.
\end{equation*}
The constant $\mathcal{N}_0\qty( \{\E{t}\} )$ which depends on the $\E
t$ only can then be fixed
by using the OPE. 
The last equation reproduces  the usual bosonization
  procedure.

  In a similar way we can compute all the correlators as
  \begin{align*}
    &\frac{\GGexcvacbra
    R\qty[ \prod\limits_i \Uppsi(x_i) \prod\limits_j \Uppsi^*(x_j) ]
    \GGexcvacket
    }{
    \braket{ \GGexcvac }
      }
\nonumber\\
&=
              \frac{
        \left\langle
          R\qty[
              \prod\limits_i \Uppsi(x_i) \prod\limits_j \Uppsi^*(x_j)
              \prodt \spin{\E{t}}{x_t}
          ]
              \right\rangle
      }{
        \left\langle
          R\qty[
          \prodt \spin{\E{t}}{x_t}
          ]
        \right\rangle
      }
,
    \end{align*}
by using Wick theorem since the algebra and the action of creators and
annhilators is the usual.
In particular taking one $\Uppsi(z)$  and one $\Uppsi^*(w)$ we get the
Green function which is nothing else but the contraction in equation
\eqref{eq:gen_Radial_order} exactly as in the usual case. 
    
  \footnotesize
  \bibliographystyle{utphys}
  \bibliography{twist_spin_branes}
  \normalsize


  \appendix

    \section{
      Details on Reflection Condition on the Vacuum
      with an Arbitrary Number of Defects
      for Asymptotic Field
    }
\label{app:details_reflection}
    We would like to provide some details on how
    \eqref{eq:reflection condition_out_field_generic_vacuum} can
    be derived.
    First we introduce the projector of positive frequency and
    negative frequency modes for the NS fermion as
    \begin{eqnarray*}
      P^{(+,0)}(z,w)
      & = &
      \frac{+1}{z-w}, \quad \abs{z} > \abs{w}
      \\
      P^{(-,0)}(z,w)
      & = &
      \frac{-1}{z-w}, \quad \abs{z} < \abs{w},
    \end{eqnarray*}
    so that for example
    \begin{equation*}
      \oint\limits_{\abs{z} > \abs{w}} \frac{\dd{w}}{ 2\uppi i}
      P^{(+,0)}(z,w)
      \Uppsi^{(0)}( 0 )
      =
      \Uppsi^{(0, +)}( z )
      ,
    \end{equation*}
    and similarly for the negative frequency modes.

    Likewise we introduce the projectors for the field with
    defects as
    \begin{eqnarray*}
      P^{(+)}(z,w)
      & = &
      \frac{P(z; \{x_t, \E t\} ) P(w; \{x_t, -\E t\} ) }{z-w},
      ~~\quad \abs{z} > \abs{w}
      \\
      P^{(-)}(z,w)
      & = &
      \frac{-P(z; \{x_t, \E t\} ) P(w; \{x_t, -\E t\} )}{z-w},
      \quad \abs{z} < \abs{w},
    \end{eqnarray*}
    with $P(z; \{x_t, \E t\} ) = \prod_{t=1}^N \qty( 1- \frac{z}{x_t} )^{\E t}
    $ as in the main text.
    
    It is then immediate to compute
    \begin{align*}
      \qty(P^{(+)} P^{(+,0)})(z,w)
      &=
        \oint\limits_{\abs{z}>\abs{\upzeta}>\abs{w}}
        \frac{d \upzeta}{ 2\uppi i}
        P^{(+)}(z,\upzeta)
      P^{(+,0)}(\upzeta,w)
        =
        P^{(+,0)}(z,w)
      &
        \nonumber\\
      \qty(P^{(+)} P^{(-,0)})(z,w)
      &=
        \frac{ P(z; \{x_t, \E t\} ) P(w; \{x_t, -\E t\} ) -1
        }{
        z-w}.
      &
    \end{align*}
The last equation is valid when $\M=\sumt \E t \le 0$ and for $\abs{z}$ and
$\abs{w}$ arbitrary.

Specializing the previous expressions to the $\Uppsi^{(out)}( z )$
case we need to add the constraints that $\abs{z} > x_1$ and $\abs{w} > x_1$.    

Finally the vacuum in presence of defects can be  described by
\begin{align*}
  \Uppsi^{(+)}( z )\Gexcvacket
  &=
    \qty(P^{(+)}\Uppsi)( z )\Gexcvacket
    \nonumber\\
  &=
    \qty(P^{(+)}\Uppsi^{(out)})( z )\Gexcvacket
    \nonumber\\
  &=
    \qty[
    \qty(P^{(+)}P^{(+,0)}\Uppsi^{(out)})( z )
    +
    \qty(P^{(+)}P^{(-,0)}\Uppsi^{(out)})( z )
    ]
    \Gexcvacket
    \nonumber\\
    &=0
    ,
\end{align*}
where we assumed $\abs{z} > x_1$ and which immediately becomes \eqref{eq:reflection condition_out_field_generic_vacuum}.

\end{document}